\def\isnotanon{1}
\documentclass[conference,compsoc]{IEEEtran}

\ifCLASSOPTIONcompsoc
  \usepackage[nocompress]{cite}
\else
  \usepackage{cite}
\fi

\usepackage{soul}
\usepackage{listings}

\usepackage{mfirstuc}

\usepackage{graphicx}
\usepackage{tabularx}
\usepackage{booktabs}
\usepackage{etoolbox}
\usepackage[nointegrals]{wasysym}

\usepackage{url}
\usepackage{hyperref}

\usepackage{flushend}

\usepackage{glossaries}
\glsdisablehyper

\usepackage{enumitem,amssymb}

\newlist{todolist}{itemize}{2}
\setlist[todolist]{label=$\square$}
\usepackage{pifont}

\usepackage[makeroom]{cancel}

\usepackage{graphicx}
\usepackage{subcaption}
\usepackage{comment}
\usepackage{xspace}

\usepackage{multirow}
\usepackage{soul}
\usepackage{tikz}
\usepackage{cleveref}

\usepackage{amsthm}
\usepackage{amsfonts}
\usepackage{mathtools}
\usepackage[utf8]{inputenc}

\usepackage{bm}

\usepackage{color}
\usepackage{xcolor}

\usepackage{enumitem}
\setlist[itemize]{noitemsep, topsep=0pt, leftmargin=10pt}
\setlist[enumerate]{noitemsep, topsep=0pt, leftmargin=10pt}

\usepackage{colortbl}
\definecolor{Gray}{gray}{0.65}
\definecolor{LightGray}{gray}{0.9}

\usepackage{array}
\usepackage{arydshln}
\newcommand{\cfref}[1]{c.f.~\S\ref{#1}}

\newcommand{\subsecspacingtop}{}%
\newcommand{\subsecspacingbot}{}%

\newcommand{\fakeparagraph}[1]{\vskip 0pt\noindent\textbf{#1.}}

\newcommand{\oursystem}{Cohere\xspace}
\newglossaryentry{privatekube}{name={PrivateKube}, description={}}

\newglossaryentry{manager}{name={budget control system}, description={}}
\newglossaryentry{planner}{name={resource planner}, description={}}
\newglossaryentry{sharespace}{name={privacy space}, plural={privacy spaces}, description={}}
\newglossaryentry{allocation}{name={allocation}, plural={allocations}, description={}}

\newglossaryentry{userid}{name={user ID}, plural={user IDs}, description={}}

\newglossaryentry{group}{name={group}, plural={groups}, description={At every allocation, a group of user is exchanged (retired/created)}}
\newglossaryentry{groupid}{name={group ID}, plural={group IDs}, description={}}

\newglossaryentry{usergroup}{name={user group}, plural={user groups}, description={}}
\newglossaryentry{userdata}{name={user data}, plural={user data}, description={}}
\newglossaryentry{blockid}{name={block ID}, plural={block IDs}, description={}}
\newglossaryentry{segment}{name={segment}, plural={segments}, description={}}
\newacronym[plural=PAs,firstplural=partitioning attributes (PAs)]{pa}{PA}{partitioning attribute}

\newacronym{dpf}{DPF}{Dominant Private Block Fairness}
\newglossaryentry{dpk}{name={DPK}, plural={DPKs}, description={}}
\newacronym{fcfs}{FCFS}{First Come, First Served}

\newacronym{ilp}{ILP}{Integer Linear Program}

\newacronym{dp}{DP}{Differential Privacy}
\newacronym{adp}{ADP}{$(\epsilon, \delta)$ - Differential Privacy}
\newacronym{rdp}{RDP}{Rényi Differential Privacy}

\newcommand\labelAndRemember[2]
  {\expandafter\gdef\csname labeled:#1\endcsname{#2}%
   \label{#1}#2}
\newcommand\recallLabel[1]
   {\csname labeled:#1\endcsname\tag{\ref{#1}}}

\newcommand{\mscale}[2]{\scalebox{#1}{$#2$}}

\newrobustcmd*{\mysquare}[1]{\tikz{\filldraw[draw=#1,fill=#1] (0,0)rectangle (0.2cm,0.2cm);}}
\newrobustcmd*{\mycircle}[1]{\tikz{\filldraw[draw=#1,fill=#1] (0,0) circle [radius=0.1cm];}}
\newrobustcmd*{\mytriangle}[1]{\tikz{\filldraw[draw=#1,fill=#1] (0,0) -- (0.2cm,0) -- (0.1cm,0.2cm);}}

\newtheorem{theorem}{Theorem}

\usepackage{lipsum}
\usepackage{mfirstuc}

\AtBeginEnvironment{appendices}{\crefalias{section}{appendix}}

\begin{document}

\title{\oursystem: Managing Differential Privacy in Large Scale Systems}

\def\isnotanon{1}

\author{{\rm Nicolas K\"uchler\textsuperscript{1}, Emanuel Opel\textsuperscript{1}, Hidde Lycklama\textsuperscript{1}, Alexander Viand\textsuperscript{2}, Anwar Hithnawi\textsuperscript{1}}  \\
\\
{\textsuperscript{1}\textit{ETH Zurich} \ \textsuperscript{2}\textit{Intel Labs}}}

\maketitle
\begin{abstract}
The need for a privacy management layer in today's systems started to manifest with the emergence of new systems for privacy-preserving analytics and privacy compliance.
As a result, many independent efforts have emerged that try to provide system support for privacy.
Recently, the scope of privacy solutions used in systems has expanded to encompass more complex techniques such as \gls{dp}.
The use of these solutions in large-scale systems imposes new challenges and requirements.
Careful planning and coordination are necessary to ensure that privacy guarantees are maintained across a wide range of heterogeneous applications and data systems.
This requires new solutions for managing and allocating scarce and non-replenishable privacy resources.
In this paper, we introduce \oursystem, a new system that simplifies the use of \gls{dp} in large-scale systems.
\oursystem implements a unified interface that allows heterogeneous applications to operate on a unified view of users' data.
In this work, we further address two pressing system challenges that arise in the context of real-world deployments: ensuring the continuity of privacy-based applications (i.e., preventing privacy budget depletion) and effectively allocating scarce shared privacy resources (i.e., budget) under complex preferences.
Our experiments show that Cohere achieves a 6.4--28x improvement in utility compared to the state-of-the-art across a range of complex workloads.
\glsreset{dp}
\end{abstract}

\IEEEpeerreviewmaketitle

\section{Introduction}
\label{sec:intro}
\noindent
Recent years have seen an exponential increase in the amount of sensitive user data that organizations collect.
However, this accumulation leaves users vulnerable to privacy violations and data abuse~\cite{Cheng2017-uj, abuse-capitalone, abuse-cambridgeanalytica}.
This has raised societal concerns and led to a shift in public perception as well as an increase in regulatory demands to protect personal data~\cite{GDPR2016-gdpr, CCPA2018-ccpa}.
Ensuring compliance with new data privacy regulations, such as the General Data Protection Regulation (GDPR) and the California Consumer Privacy Act (CCPA), introduces a variety of challenges for today's data systems~\cite{Shastri2020-gdprbench, Shah2019-nj, Shastri2019-gdprsins}.
Additionally, a shift in general expectations of digital privacy has induced a demand for privacy-enhancing solutions that can protect sensitive data while retaining its value.
The adoption of these solutions into current systems introduces further unique challenges.
Privacy requirements sometimes fundamentally conflict with conventional systems' design goals such as optimizing for availability, performance, and scalability.
This further complicates the integration of privacy solutions into existing systems~\cite{Shastri2020-gdprbench}.
The complexity of managing and integrating privacy into existing large-scale systems has spurred the development of numerous frameworks and tools that simplify both compliance with privacy regulations and the development of privacy-focused applications.
More generally, a new domain of `Privacy Engineering' is emerging to answer these new requirements.
Tools such as data tagging for personally identifiable information (PII)~\cite{Logozzo2020-tagging}, data cataloging and lineage~\cite{Databricks2021-unity}, or privacy transformations~\cite{Privitar2016-web}, have made it easier for organizations to enforce compliance and develop privacy-focused technologies and practices.
Recently, the scope of privacy-enhancing solutions has expanded alongside a demand for more rigorous privacy risk management.

\noindent\Gls{dp}~\cite{Dwork2006-originaldp} is widely recognized as the gold standard for protecting personal information and provides a mathematically rigorous definition of privacy.
It allows data to be processed and released while minimizing the risk of re-identification for individual users.
\Gls{dp} is rapidly evolving from academic theory to real-world deployment in large-scale systems~\cite{Desfontaines2022-dppractice, Aktay2020-dpgooglecovid3, Bavadekar2020-dpgooglecovid2, Bavadekar2021-dpgooglecovid, Rogers2020-dplinkedinlabor}.
Today, \gls{dp} is widely used in a variety of settings, including the US Census~\cite{Census2022-web}, and in user data collection at Apple~\cite{Apple2017-dp}.
The use of \gls{dp} is expected to grow even further~\cite{Gartner2022-privacyhype, Desfontaines2022-dpubiquitous}.
While there has been significant work on the theoretical and technical aspects of \gls{dp}, scant attention has been paid to the practical challenges of deploying \gls{dp}-based solutions in complex, real-world systems.

\vspace{0.2em}
\fakeparagraph{Differential Privacy Systems Challenges}
The complexity of using \gls{dp} has led to a flurry of frameworks and libraries that focus on integrating \gls{dp} into existing data frameworks and algorithms.
Examples include TensorFlow Privacy~\cite{Erlingsson2019-tfprivacy}, Tumult Analytics~\cite{Tumult2022-web}, and DP-SGD~\cite{Abadi2016-mldp}.
However, these focus primarily on improving the development experience for individual applications and ignore the complexities and challenges that arise when using \gls{dp} in larger systems.
In large-scale systems, careful planning and coordination are required to ensure that privacy guarantees are maintained across a wide range of heterogeneous applications and data systems.
This raises new challenges in managing shared, scarce, and non-replenishable differentially private resources.
Without careful management, \gls{dp} can quickly become excessively restrictive.
As \gls{dp} matures and its adoption continues to increase, these deployment issues are rapidly moving to the foreground.

The community has recently acknowledged the crucial role of system infrastructure for \gls{dp} in driving its adoption~\cite{Cummings2023-dpfrontier}.
However, existing frameworks for \gls{dp} fall short in this, as they generally only offer support for individual (types of) applications.
While we can deploy multiple such frameworks in parallel, this requires making non-collusion assumptions that might be hard to justify in practice.
Recent work by Luo et al.~\cite{Luo2021-privacysched} was the first to recognize the need for managing privacy resources for workloads of machine learning pipelines in a unified way.
They proposed a dedicated privacy resource scheduling algorithm that operates alongside a cluster hardware scheduler (i.e., Kubernetes).
While this is an important first step in terms of addressing the lack of privacy management tailored to \gls{dp} in current systems, their approach does not generalize well to broader organizational privacy management needs.
In order to address these needs, we need to consider large-scale systems featuring a wide variety of independent applications.
These must be coordinated and controlled in a centralized and unified manner to ensure the privacy of the entire system.
Organizations require proactive control of how their privacy resources are distributed and utilized, taking into account applications' relative priorities and ensuring that they can be easily integrated into existing data architectures.

In this paper, we present \oursystem, a privacy management system designed to facilitate the operation of \gls{dp} solutions in large-scale systems.
\oursystem addresses the above challenges via a new end-to-end DP platform design.
Concretely, our work focuses on addressing the following challenges:

\vspace{0.1em}
\fakeparagraph{{\large \textcircled{\normalsize 1}} Unified System Architecture for Privacy}
Achieving a global \gls{dp} guarantee across multiple systems necessitates a comprehensive view of data with consistent and accurate user association.
Fortunately, other essential aspects of modern data systems, such as Data Governance, Compliance, and Data Discovery, also share these requirements.
As a result, there has been an increasing interest in data management architectures that enable the discovery, tracking, and auditing of heterogeneous data in large-scale systems.
These approaches frequently build upon data catalog systems~\cite{Google2022-datacatalog, Microsoft2022-datacatalog, Databricks2021-unity, IBM2022-datacatalog} that act as centralized metadata repositories documenting information about the available data assets within an organization.
The unified global view they offer makes them ideal candidates for integrating security and privacy controls.
Most data catalog systems already support basic access control and encryption~\cite{Google2022-datacatalog, Microsoft2022-datacatalog, IBM2022-datacatalog}, and there are efforts to extend their functionality with more fine-grained access control~\cite{Immuta2021-web, Ethyca2021-web, Databricks2021-unity}, data governance tools~\cite{Ethyca2021-web, Privacera2022-web}, and basic privacy transformations~\cite{Privitar2016-web, Privacera2022-web, Immuta2021-web} (e.g., data minimization).
These systems can naturally support privacy techniques which can be achieved through pre-processing data to generate privacy-compliant views.
\Gls{dp} mechanisms, however, cannot be accommodated purely through pre-processing.
For example, modern differentially private machine learning training need to access raw data, with noise only being added to the model incrementally during training~\cite{Abadi2016-mldp}.
As a result, \gls{dp} management requires new approaches beyond simple view-based access control.
Moreover, \gls{dp} guarantees do not exist in isolation but depend on the whole history of \gls{dp} applications applied to the data.
Existing systems lack the ability to account for, and track, this shared global state and therefore cannot be easily extended with support for DP management.
\oursystem implements a unified abstraction that allows applications to generically express their privacy resource needs and allows a variety of \gls{dp} systems to operate on a shared collection of data using a wide range of techniques.
Applications remain agnostic of how data and privacy resources are shared and managed; hence allowing multiple private systems to operate on the same data; akin to how conventional (i.e., non-\gls{dp}) applications operate today.

\vspace{0.2em}
\fakeparagraph{{\large \textcircled{\normalsize 2}} Effective Use of Privacy Resources}
Developing \gls{dp} applications is a challenging task that requires careful consideration of algorithm design, privacy analysis, and hyperparameters to provide meaningful privacy guarantees while maintaining acceptable utility and fairness~\cite{Miklau2022-dpcasesalary, Papernot2021-dphyperparam}.
Furthermore, when moving to the multiple applications setting with diverse mechanisms, compatibility issues between privacy analysis tools arise, adding another layer of complexity.
More specifically, determining an accurate bound on the privacy impact of many different DP mechanisms is a key challenge to the effective utilization of DP in large-scale real-world scenarios.
Specifically, generic (approximate \gls{dp}) composition~\cite{Dwork2010-dpadcomp, Kairouz2017-dpoptcomp} generally introduces a significant gap between the actual privacy cost and the bounds they provide.
While the DP literature has introduced many optimized accounting methods to improve the privacy analysis of individual applications~\cite{Feldman2020-renyifilter, Abadi2016-mldp, Rogers2021-dpboundedrange, Dong2019-dpboundedrange}, most techniques do not apply to settings with mixed workloads of different DP applications.

Beyond the compatibility of privacy analysis tools, there is a more fundamental issue, when deploying multiple \gls{dp} applications.
Traditionally, the analysis of a single application has considered all data to be part of a single fixed dataset and then examined the impact of each query on the global privacy budget.
However, real-world applications frequently consider subsets of users defined by their characteristics, which state-of-the-art DP approaches can leverage to reduce the privacy cost of such queries even when the characteristics themselves (e.g., age, or region) are privacy sensitive.
In \oursystem, we extend the privacy management with native support for such \emph{partitioning} attributes, so that the privacy management layer can fully realize these privacy cost savings.
However, a more detailed privacy analysis increases the management complexity, as it requires keeping track of additional metadata about data usage.
In this paper, we show how to efficiently track fine-grained privacy budgets without the need to explicitly track all possible subsets (\cfref{sec:dp:pa}).
We further show how to integrate this approach with the \emph{privacy amplification via subsampling} technique, which \oursystem utilizes to improve the privacy costs of queries over random subsets of the users.
In combination, this enables \oursystem to significantly improve the privacy analysis of varied workloads resulting in more utility under the same privacy budget, compared to the state-of-the-art~\cite{Luo2021-privacysched}.

\fakeparagraph{{\large \textcircled{\normalsize 3}} Sustaining Systems Continuity under DP}
Addressing the need of real-world systems to perform continuous operations on data is one of the most pressing challenges in real-world \gls{dp} deployment.
In this setting, the available (finite) privacy budget is bound to be depleted over time even with the most accurate privacy accounting.
In practical deployments, this issue is often circumvented by periodically refreshing the privacy budget of users~\cite{Apple2017-dp}, however, this fundamentally undermines the guarantees of DP.
Instead, we can exploit the fact that, in large real-world systems, there is virtually always a steady influx of new users.
While privacy budgets for individual users must be finite (and comparatively small, in order to achieve strong guarantees), we can retire users and replace them with new users to replenish the overall available budget.
However, this user rotation needs to be carefully implemented to minimize the need for additional tracking, maintain fairness over time, and provide clear semantics for DP applications.
We show how to augment sliding-window-based user rotation with user activation and budget-unlocking strategies in order to achieve these properties.
By combining these techniques, our system presents the first practical and comprehensive solution for managing privacy in the context of real-world, continuous, and diverse DP workloads.

\fakeparagraph{\oursystem}
We design and implement \oursystem, a privacy management system for large-scale deployments of \gls{dp}. We evaluate our system on a diverse range of workloads, demonstrating its ability to handle a complex mix of applications and \gls{dp} mechanisms.
Our evaluation results show that \oursystem improves upon the state-of-the-art significantly, providing a 6.4--28x improvement in utility.
As part of this work, we also develop an evaluation framework and workload generator to evaluate large-scale mixed DP application deployments.
We make \oursystem, alongside our workload generator, available as open-source resources.

\subsecspacingtop
\section{Preliminaries}
\subsecspacingbot
\label{sec:bg}

\subsection{Threat Model}
\label{sec:bg:threat}
\noindent
We consider the global model of \gls{dp}~\cite{Desfontaines2019-dpgloballocal}, where the party computing the \gls{dp} mechanisms has access to the underlying data.
We note that \gls{dp}, by definition, holds against an unconstrained adversary with arbitrary computational power and background knowledge.
We assume the individual \gls{dp} mechanisms are implemented correctly, e.g., they do not contain floating point vulnerabilities~\cite{Mironov2012-dpfloating, Haney2022-dpfloating}, under-estimate the required noise, or otherwise violate their stated privacy guarantees.
Lastly, we assume that there is no (side-channel) leakage~\cite{Haeberlen2011-dpfire} beyond the intentional data releases.

\subsection{Differential Privacy}
\label{sec:bg:dp}
\noindent
\gls{dp} is considered the gold standard for the privacy-preserving release of data.
In essence, a \gls{dp} mechanism hides the contribution of an individual by introducing carefully calibrated randomness during the data analysis.
More formally, a randomized algorithm $\mathcal{M}$ is $(\epsilon, \delta)$-differentially private if for any set of results $S \subseteq Range(\mathcal{M})$ and any two neighboring datasets $D, D'$ which only differ in a single individual, $Pr[\mathcal{M}(D)\in \mathcal{S}] \leq exp(\epsilon) \cdot Pr[\mathcal{M}(D') \in \mathcal{S}] + \delta$.
Two datasets are neighboring if $D'$ can be constructed by adding or removing one single individual from $D$.
The privacy parameters $\epsilon > 0$ and $\delta \in [0, 1)$ quantify the privacy cost of a data release.
In addition to this definition, known as \gls{adp}, there are different notions of \gls{dp} that analyze it using slightly different privacy parameters.
The maximum privacy parameters (e.g., $\epsilon, \delta$) that one is willing to tolerate, are frequently referred to as the \emph{privacy budget}.
A wide variety of \gls{dp} mechanisms allows computing aggregate statistics~\cite{Erlingsson2014-rappor, Wilson2020-dpsqlgoogle}, training ML models~\cite{Abadi2016-mldp}, or other applications~\cite{Mueller2022-dpgraphsok, Roth2021-mycelium, Roth2020-orchard, Roth2019-honeycrisp}.
The majority of mechanisms are based on adding noise.
For example, the Gaussian Mechanism takes a function $f: D \rightarrow \mathbb{R}$ with sensitivity $\Delta_f$ and outputs $\mathcal{M}(X) = f(x) + \mathcal{N}(0, \sigma^2)$ where $\sigma^2 = 2\Delta_f^2 \log{(1.25/\delta)}/\epsilon^2$ satisfies $(\epsilon, \delta)$-DP~\cite{Dwork2006-dpgaussian}.

\subsection{Differential Privacy Composition}
\label{sec:bg:dp:comp}
\noindent
\Gls{dp} can formalize the cumulative privacy loss of multiple data releases, even when using different mechanisms.
We briefly introduce three key types of composition and describe our setting and assumptions.

\noindent
\textbf{Parallel composition} states that if we partition the dataset into $m$ disjoint datasets, and then run an $(\epsilon_i, \delta_i)$ differentially private mechanism $M_i$ on each partition individually, the joint privacy loss is $max_{i} (\epsilon_i, \delta_i)$.
This formalizes the idea that releasing data on one set of users does not impact the \gls{dp} guarantees for other users.

\noindent
\textbf{Sequential Composition} states that for a fixed series of $(\epsilon_i, \delta_i)$ differentially private mechanisms $M_i$, the cumulative privacy loss is bounded by $(\sum_i \epsilon_i, \sum_i \delta_i)$.
This bound can be made tighter in practice through the use of more advanced composition theorems~\cite{Dwork2010-dpadcomp, Kairouz2017-dpoptcomp, Whitehouse2022-dpfilterod}.

\fakeparagraph{Fully Adaptive Concurrent Setting}
The work on \gls{dp} composition distinguishes between settings where the series of mechanisms and their privacy parameters are known a priori or can be chosen adaptively.
\emph{Fully adaptive concurrent composition}~\cite{Haney2023-dpfadaptconc} is the most flexible setting that permits mechanisms to use previous results and adapt the choice of their privacy parameters based on them.
Moreover, in this setting, queries can be arbitrarily interleaved among different concurrent mechanisms, rather than being restricted to a sequential order.
Frequently, the goal is to ensure that a series of adaptively chosen mechanisms does not exceed a pre-specified budget, instead of bounding the privacy parameters afterward.
This is achieved by a \emph{privacy filter}~\cite{Rogers2016-dpodometer, Whitehouse2022-dpfilterod}, which provides a stopping rule that prevents mechanisms from exceeding a fixed budget.

\subsection{Rényi Differential Privacy}
\label{sec:bg:dp:rdp}
\noindent
In recent years, several variants of \gls{dp} have emerged that have more appealing composition properties than traditional \gls{adp}.
This includes \gls{rdp}~\cite{Mironov2017-rdp}, which has gained increasing popularity.
A randomized mechanism $M$ satisfies $(\alpha,\epsilon^{(rdp)})$ \gls{rdp} if:
$D_\alpha(M(D) \lVert M(D')) \leq \epsilon^{(rdp)}$, where $D_\alpha$ is the Rényi Divergence of order $\alpha$ for $\alpha > 1$.
\gls{rdp} has equivalent composition theorems for parallel,  sequential, and concurrent composition~\cite{Mironov2017-rdp, Haney2023-dpfadaptconc}.
Specifically, for a series of $(\alpha, \epsilon^{(rdp)}_i)$ differentially private mechanisms, the composition satisfies $(\alpha, \sum \epsilon^{(rdp)}_i)$.
\gls{rdp} typically allows for much tighter compositions, in particular when using the Gaussian mechanism.
\gls{rdp} privacy costs can be converted back to \gls{adp}, and
$(\alpha, \epsilon^{(rdp)})$-\gls{rdp} implies $(\epsilon^{(rdp)}+\frac{\log 1 / \delta}{\alpha - 1}, \delta)$-\gls{adp}.\!\footnote{The other direction does not hold, i.e., \gls{rdp} is strictly stronger.}

\fakeparagraph{Budget and Privacy Filter}
From a practical perspective, the above two properties of \gls{rdp} composition allow tracking not only a single order $\alpha$ but a set of orders $\{ \alpha_a \}$.
Given a fixed $\delta$, the best achievable $\epsilon$ can then be determined by checking the conversion for every order $\alpha$.
Since the optimal order $\alpha$, i.e., which yields the best $(\epsilon, \delta)$-budget, is contingent upon the particular mechanisms employed in the composition,  it is not possible to determine it beforehand.
Nonetheless, by monitoring a limited number of orders, a tight privacy analysis can be attained~\cite{Mironov2017-rdp}.
Tracking this set of values allows building a simple privacy filter~\cite{Lecuyer2021-dpodometers, Feldman2020-renyifilter} with an $(\epsilon, \delta)$-budget.
Before answering a query, the filter checks whether there exists an order $\alpha$ for which the  $(\epsilon, \delta)$-budget is satisfied and only then answers the query.

\subsecspacingtop
\section{ \oursystem Architecture}
\begin{figure}[t]
	\center
	\includegraphics[width=\columnwidth]{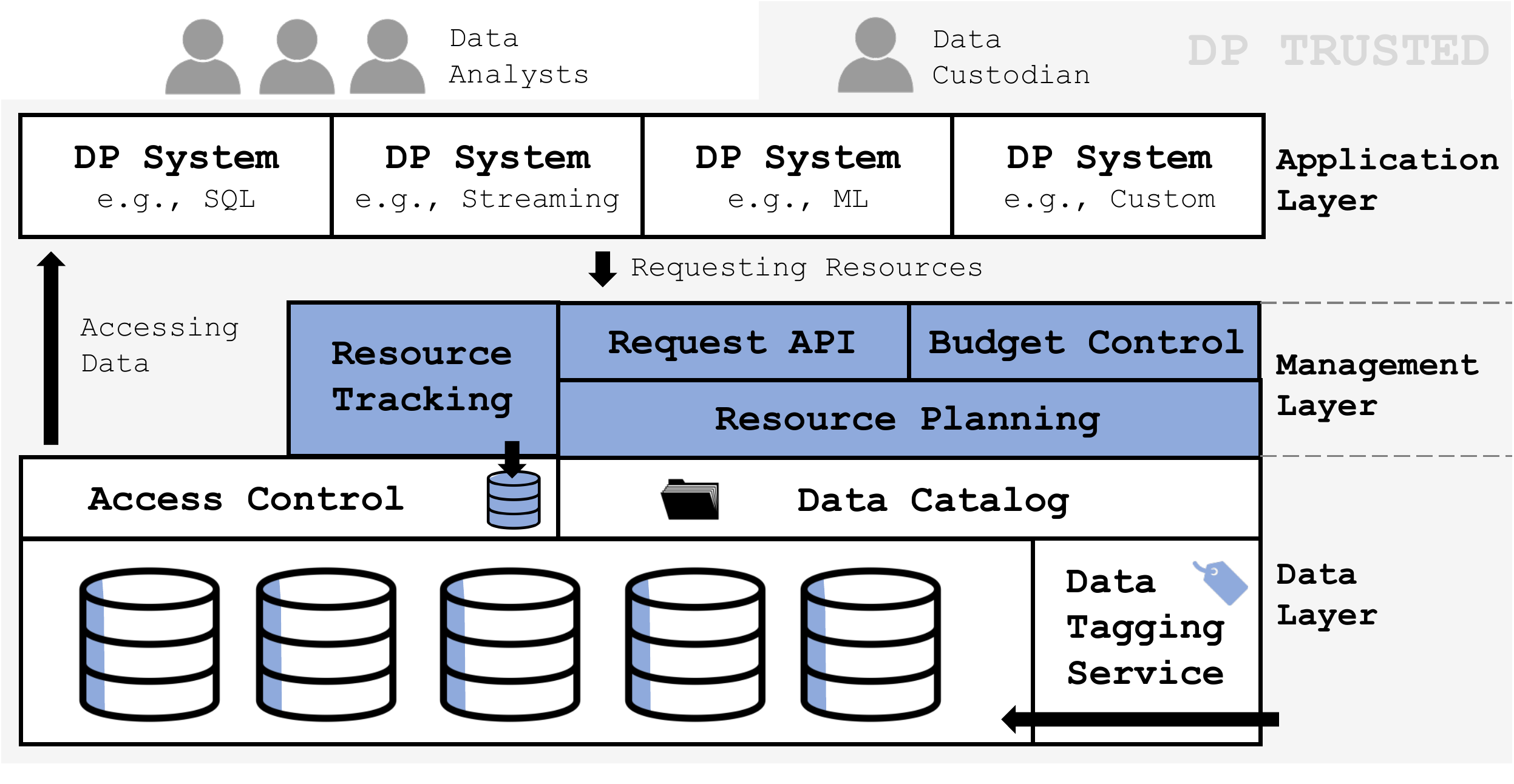}
	\caption{
		\oursystem's Architecture.
	}
	\label{fig:arch}
\end{figure}

\subsecspacingbot
\label{sec:arch}
\noindent
\oursystem, shown in \Cref{fig:arch}, is a new privacy management architecture designed to facilitate the operation of \gls{dp} solutions in large-scale systems.
To enable the enforcement of system-wide \gls{dp} privacy guarantees and effectively allocate privacy resources, a global view of data and its usage is required.
Existing work proposes to co-locate privacy management with hardware resource allocation, viewing privacy budgets as special resources.
However, hardware allocation systems do not necessarily provide a global view (i.e., are frequently deployed per cluster).
We instead identify the emerging Lakehouse architecture as a more suitable basis for privacy management.
Lakehouse architectures~\cite{Armbrust2020-deltalake, ApacheIceberg2021-web, ApacheHudi2021-web} add additional metadata that facilitates a centralized data view and provides a foundation for conventional data management (e.g., access control).
In \oursystem, we build upon this paradigm and introduce a new \gls{dp} management layer on top of this foundation.
Specifically, \oursystem's architecture comprises \emph{(i)}~an extension of the data layer with privacy-specific metadata, \emph{(ii)}~an application layer API that allows \gls{dp} applications to express their privacy needs, and \emph{(iii)}~a privacy management layer providing tools to optimally distribute and manage shared privacy resources.

\subsecspacingtop
\subsection{Data Layer}
\subsecspacingbot
\label{sec:arch:data}
\noindent
In our design, we retain the existing data layer (i.e., storage layer) and extend it with privacy-specific metadata needed for privacy analysis and enforcement.

\fakeparagraph{Data Tagging}
To support fine-grained privacy analysis, we require
the means to precisely identify the user associated with a particular data item.
In \oursystem, each user is represented by a globally unique \emph{\gls{userid}}, and we rely on a data tagging service that tags all \gls{userdata} with the \gls{userid} at ingest~\cite{Bhajaria2021-dataprivacy}.
Such services are already present in many modern large-scale systems to ensure regulatory compliance (e.g., with the GDPR).
In addition to tagging data items, we also introduce \emph{\glspl{pa}} $\texttt{A}_\texttt{1}$,\ldots, $\texttt{A}_\texttt{k}$ over the users (e.g., year of birth, region) and a grouping of users with \glspl{groupid}.
Utilizing these attributes, we define a \emph{\gls{blockid}} as a tuple (\texttt{\gls{groupid}}, $\texttt{a}_\texttt{1}$, \ldots, $\texttt{a}_\texttt{k}$).
For example, (\texttt{group51, 1975, \ldots, USA}).
These \glspl{blockid} serve as the basis for our fine-grained privacy analysis (\cfref{sec:dp:pa}).

Each \gls{group} consists of a set of users and  $\left|\texttt{A}_\texttt{1}\right| \times \cdots \times \left|\texttt{A}_\texttt{k}\right|$ different \glspl{blockid}.
We store the corresponding \gls{blockid} for each user in a dedicated \texttt{UserAttribute} table, which enables the system to filter users by predicates over the \glspl{pa}.
In \Cref{sec:dp:pa}, we demonstrate utilizing these fine-grained \glspl{pa} to improve the privacy analysis for applications targeting overlapping user subsets without needing to track individual \mbox{\glspl{blockid}}.

\begin{figure}[t]
	\begin{subfigure}{.49\columnwidth}
		\includegraphics[width=\linewidth]{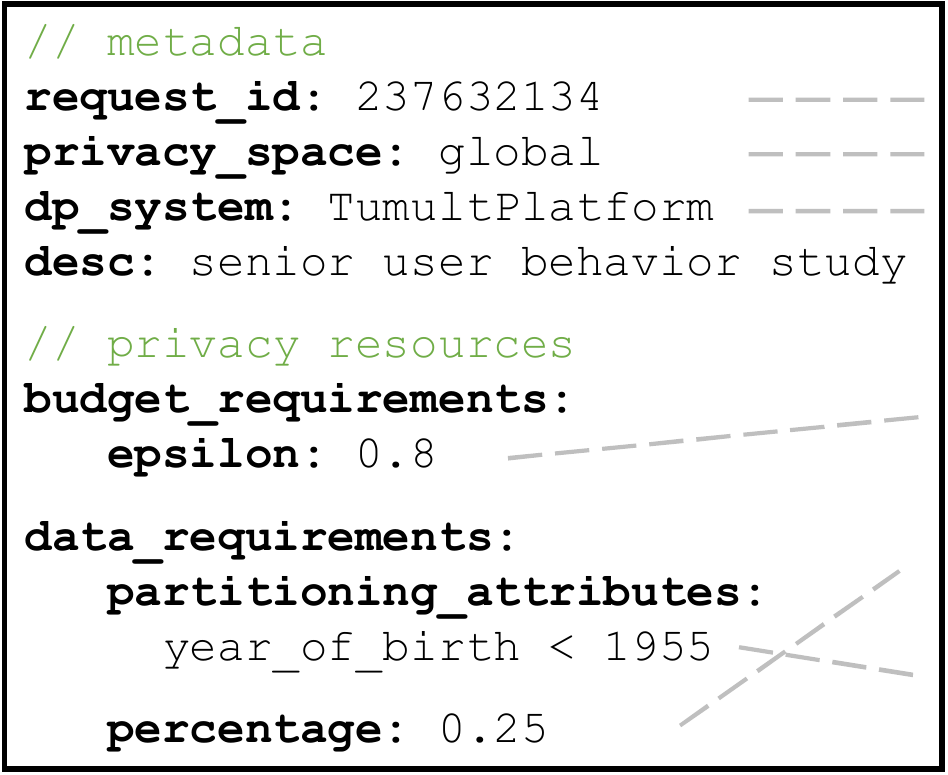}
		\caption{Data Access Request}
		\label{fig:request}
	\end{subfigure}\hfill
	\begin{subfigure}{.49\columnwidth}
		\includegraphics[width=\linewidth]{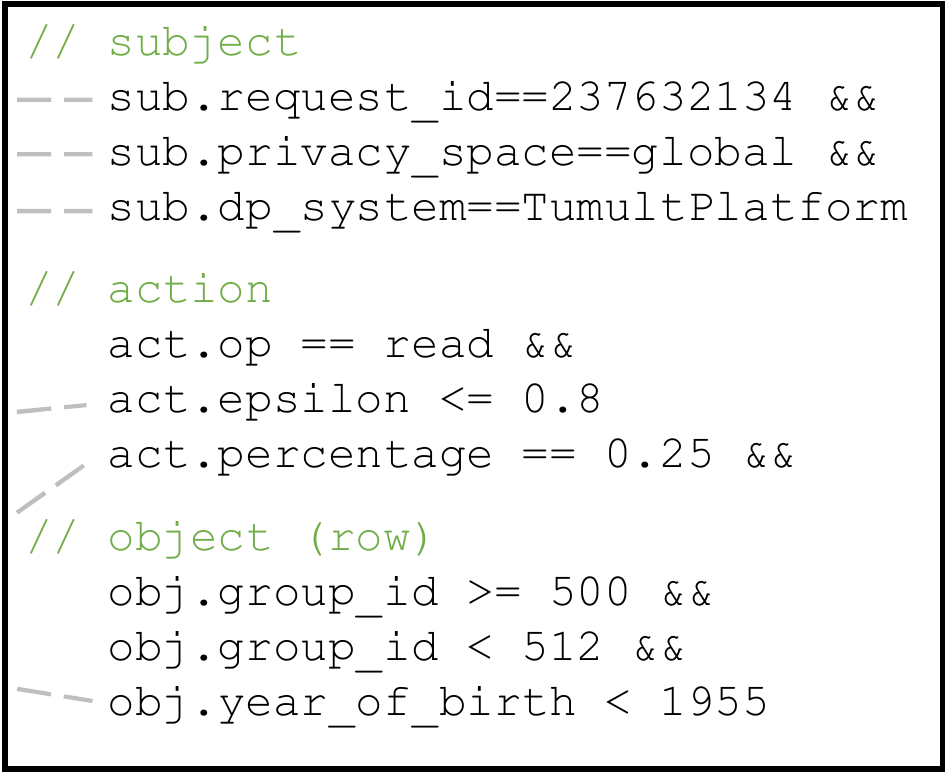}
		\caption{ABAC Policy}
		\label{fig:abac}
	\end{subfigure}
	\caption{In \oursystem, the accepted data access request on the left (a) corresponds to the ABAC policy on the right (b).}
	\label{fig:map-request-abac}
\end{figure}

\fakeparagraph{Attribute-Based Access Control}
In order to ensure that data flows in the system are compliant with the decisions made by \oursystem's privacy management layer, we employ access control at the level of \glspl{blockid}.
We use Attribute-Based Access Control (ABAC) in order to express access policies without having to explicitly enumerate large numbers of \glspl{blockid}.
ABAC is already widely deployed in practice, and \oursystem can be realized using any data layer that supports row-level ABAC.
An ABAC policy specifies a combination of attributes that is necessary for a subject to be authorized to access a resource.
In \Cref{fig:abac}, we show how our design represents a decision made by the privacy management layer as an ABAC policy.
The attributes of a subject identify an application and a \gls{dp} system, the resource attributes identify the \glspl{groupid} and optionally a number of \glspl{pa}, and the action attributes encode the \gls{dp} privacy budget together with subsampling instructions.
The set of ABAC policies defines permitted data flows and implicitly records the history of data usage.

\subsecspacingtop
\subsection{Application Layer}
\subsecspacingbot
\label{sec:arch:app}
\noindent
\oursystem exposes an application-layer API where a variety of differentially private systems can express their privacy resource needs.
\oursystem allows all of those \gls{dp} systems to operate on a unified view of the data while remaining agnostic of how data and privacy resources are shared and managed.
These \gls{dp} systems can be either existing off-the-shelf \gls{dp} solutions~\cite{Wilson2020-dpsqlgoogle, Rogers2021-dplinkedin, McSherry2009-pinq, Gaboardi2016-dppsi} targeting data analysts, or custom-built applications using \gls{dp} libraries~\cite{Yousefpour2021-opacus, Erlingsson2019-tfprivacy, Tumult2022-web, PipelineDP2022-web, OpenDP2020-whitepaper} designed for data custodians.
Today, such \gls{dp} applications are usually deployed as ad-hoc solutions operating on a fixed dataset, isolated from other applications.
\oursystem unifies these isolated \gls{dp} applications and enables them to operate on the same underlying data without violating \gls{dp} assumptions and guarantees.
Towards this, we abstract \gls{dp} applications to their data requirements and privacy costs, allowing us to support the wide variety of existing and potential future \gls{dp} applications.

\fakeparagraph{Request Data API}
Today, analysts in large organizations often use data catalogs to explore data and request access to tables and columns.
To enable support for \gls{dp} applications in this workflow, \oursystem exposes a request API to applications that allows (i) expressing precise \emph{data requirements} to enable more fine-grained resource management (\cfref{sec:arch:data}) and (ii) specifying \gls{dp} \emph{budget requirements}.
In addition, a request contains a section of general metadata.
In \Cref{fig:request}, we show an example of a request for privacy resources.
Requests can only be granted by the privacy management layer if both requirements are satisfied, as access to less data or the need to increase the noise to achieve a lower privacy cost would invalidate the utility of the application.

Specifically, the \emph{data requirements} define a subpopulation of users via a \gls{pa} filter and a percentage for subsampling.
For \emph{budget requirements}, we support expressing privacy costs in $\epsilon$-DP~\cite{Dwork2006-originaldp}, zero-Concentrated Differential Privacy ($\rho$-zCDP)~\cite{Bun2016-zCDP}, or \gls{rdp}~\cite{Mironov2017-rdp}.
Internally, all privacy costs are converted to \gls{rdp} using known conversions~\cite{Mironov2017-rdp, Bun2016-zCDP}, allowing us to use \gls{rdp} composition across all applications.
This simplifies the integration of existing \gls{dp} systems and libraries, which use a variety of different notions.
In addition, we offer an API that enables \gls{dp} libraries to expose their computation as fundamental DP mechanisms directly.
In our prototype implementation of \oursystem, we have integrated Tumult Analytics~\cite{Tumult2022-web} for SQL-like queries and the \gls{dp} version of PyTorch~\cite{Yousefpour2021-opacus} to express resource requests seamlessly within \oursystem.

\fakeparagraph{Accessing Data}
In traditional data management systems, analysts can use data processing frameworks such as Spark or Flink to connect to data they have been granted access to and interactively build  applications~\cite{Deltalake2022-integrations}.
\oursystem can preserve this workflow even in the context of \gls{dp} since our ABAC-based row-level access control inherently enforces compliance with the decisions made by the privacy management layer.
However, \gls{dp} marks a shift in access control for privacy.
Prior techniques involved pre-processing data and restricting access to filtered views, but \gls{dp} is a characteristic of the algorithm, not the data.
Integrating every possible \gls{dp} algorithm into the privacy management layer is infeasible and does not align with the separation of concerns principle.
\oursystem adopts a modular approach assigning access rights to raw data along with a budget, while the authorized \gls{dp} system ensures the privacy budget is adhered to.
This requires correct \gls{dp} system implementation, which can be ensured through auditing or privacy budget enforcement mechanisms based on trusted hardware~\cite{Wang2022-privguard}.
Note, not every new application requires a new \gls{dp} system, as a single system can be deployed to expose a safe interface for analysts to run \gls{dp} queries.
Thus, most analysts do not need to be trusted to implement \gls{dp} systems.

\subsecspacingtop
\subsection{The Privacy Management Layer}
\subsecspacingbot
\label{sec:arch:mgmt}
\noindent
The management layer in \oursystem consists of two entities: the \emph{\gls{manager}} and the \emph{\gls{planner}}.
\fakeparagraph{\Gls{manager}}
Data catalogs built for modern data ecosystems offer a global policy interface to govern and manage data resources~\cite{Immuta2021-web, Databricks2021-unity}.
In this interface, data stewards can express policies that control applications' access to data.
In \oursystem, we provide the \emph{\gls{manager}} as an equivalent system to manage \gls{dp} resources.
We support the concurrent use of a variety of \gls{dp} notions and privacy budgets to support real-world use cases occurring in organizations.
For example, each team in an organization might have an independent privacy budget for their analytics.
Such concessions are currently still frequently used to deploy \gls{dp} in practice.
\oursystem supports such use cases through \emph{\glspl{sharespace}}, a new abstraction that groups applications sharing a notion of \gls{dp} (e.g. user- or user-time-\gls{dp}\footnote{Extending \oursystem to support user-time-\gls{dp}, would require the \gls{sharespace} definition to specify how the budget is refreshed.}~\cite{Luo2021-privacysched}) and a space-specific privacy budget\footnote{In our privacy analysis, we assume that different \glspl{sharespace} are non-colluding.}.
Deploying multiple \glspl{sharespace} requires non-collusion assumptions, which might not hold in practice.
At a minimum, traditional access control systems should be used to ensure members of one space do not have direct access to releases generated in another space.
We recommend relying on this feature primarily as part of a migration strategy toward global budget accounting.
Note that even when multiple \glspl{sharespace} are used, one can still draw conclusions about the overall budget use.
In the following, we assume (without loss of generality) a `pure' setting with a single \gls{sharespace}, where all applications share the same global privacy budget.

\fakeparagraph{\Gls{planner}}
In \oursystem, the applications in each \gls{sharespace} use the RequestAPI to formulate resource requests.
The \emph{\gls{planner}} uses this information to allocate data and privacy budget to requests, creating a new allocation and associated access control policies for the data layer.
The \gls{planner} allocates the available privacy budget among (some of) the different candidate requests, taking into account their relative priorities.
Each request can be assigned a weight to express the importance of the task it is associated with.
This allows the \gls{planner} to give higher priority to requests that are critical for the organization's business needs while providing lower precedence to non-essential tasks such as supplemental analytics.
However, beyond the priority of each request, the \gls{planner} needs to consider a variety of other aspects, including the available privacy budget, and the cost of each request.
In the following section, we describe our fine-grained privacy analysis and how the \gls{planner} derives an allocation.

\section{Differential Privacy in \oursystem}
\label{sec:dp}
\noindent
In order to realize an effective DP management system, we need a fine-grained privacy analysis that determines a tight bound on the consumed privacy budget for real-world mixed workloads.
However, even with an optimal privacy analysis, a continuously running system will inevitably deplete the entire budget.
Consequently, there is a pressing need for additional techniques that effectively sustain the system's continuity.
Finally, we need an effective allocation mechanism that carefully balances the utility and cost of different requests so that the final system not only preserves privacy but also produces meaningful results.
In this section, we elaborate on how we tackle these challenges.

\subsection{Fine-grained Privacy Analysis}
\label{sec:dp:pa}
\label{sec:dp:sub}
\noindent
While the privacy costs of individual DP mechanisms are easy to derive, determining an accurate bound on the privacy impact of a multitude of DP mechanisms is more challenging.
Specifically, both DP's basic and advanced composition theorems~\cite{Dwork2010-dpadcomp} generally introduce a significant gap between the actual privacy cost and the bounds they provide.
While the DP literature has introduced many optimized accounting methods to improve the privacy analysis of individual applications~\cite{Feldman2020-renyifilter, Abadi2016-mldp, Rogers2021-dpboundedrange, Dong2019-dpboundedrange}, most techniques do not apply to settings with mixed workloads of different DP applications.
In this setting, \gls{rdp} provides the ability to more efficiently analyze a varied set of mechanisms without relying on mechanism-specific approaches by capturing more information about the underlying distributions than traditional $\epsilon$-DP or $(\epsilon,\delta)$-DP.
We base our privacy analysis on an RDP-based privacy filter (\cfref{sec:bg:dp}) and combine it with partitioning attributes and subsampling, which also match well with the mixed workload setting.
As a result, \oursystem's privacy analysis significantly outperforms prior work on privacy management systems~\cite{Luo2021-privacysched}.

\fakeparagraph{Partitioning Attributes}
In \oursystem, we improve the privacy analysis by considering that many applications target only a subset of the users (e.g., only users of a certain age or from a certain region).
As frequently noted in the literature on \gls{dp} applications~\cite{Lecuyer2019-sage, McSherry2009-pinq, Smith2022-dpparallel}, this presents an opportunity for a tighter privacy analysis by tracking privacy budgets for these subsets separately.
Intuitively, if a user's data is not influencing the outcome of an application, then this user also does not incur any privacy cost. %
Specifically, we can employ block composition~\cite{Lecuyer2019-sage} for queries on different subsets.
For example, two applications that only consider either users from region A or region B (where A and B are disjoint) need not consume the sum of their privacy costs but only the maximum of either (by parallel composition \cfref{sec:bg:dp:comp}).
Note that applying block composition requires the partitioning scheme (e.g. the set of regions) to be publicly known.
However, the resulting partitions can remain private (e.g., what region a user belongs to)~\cite{Lecuyer2019-sage}.

\begin{figure}[t]
    \begin{subfigure}[t]{.49\columnwidth}
        \includegraphics[height=5.1cm]{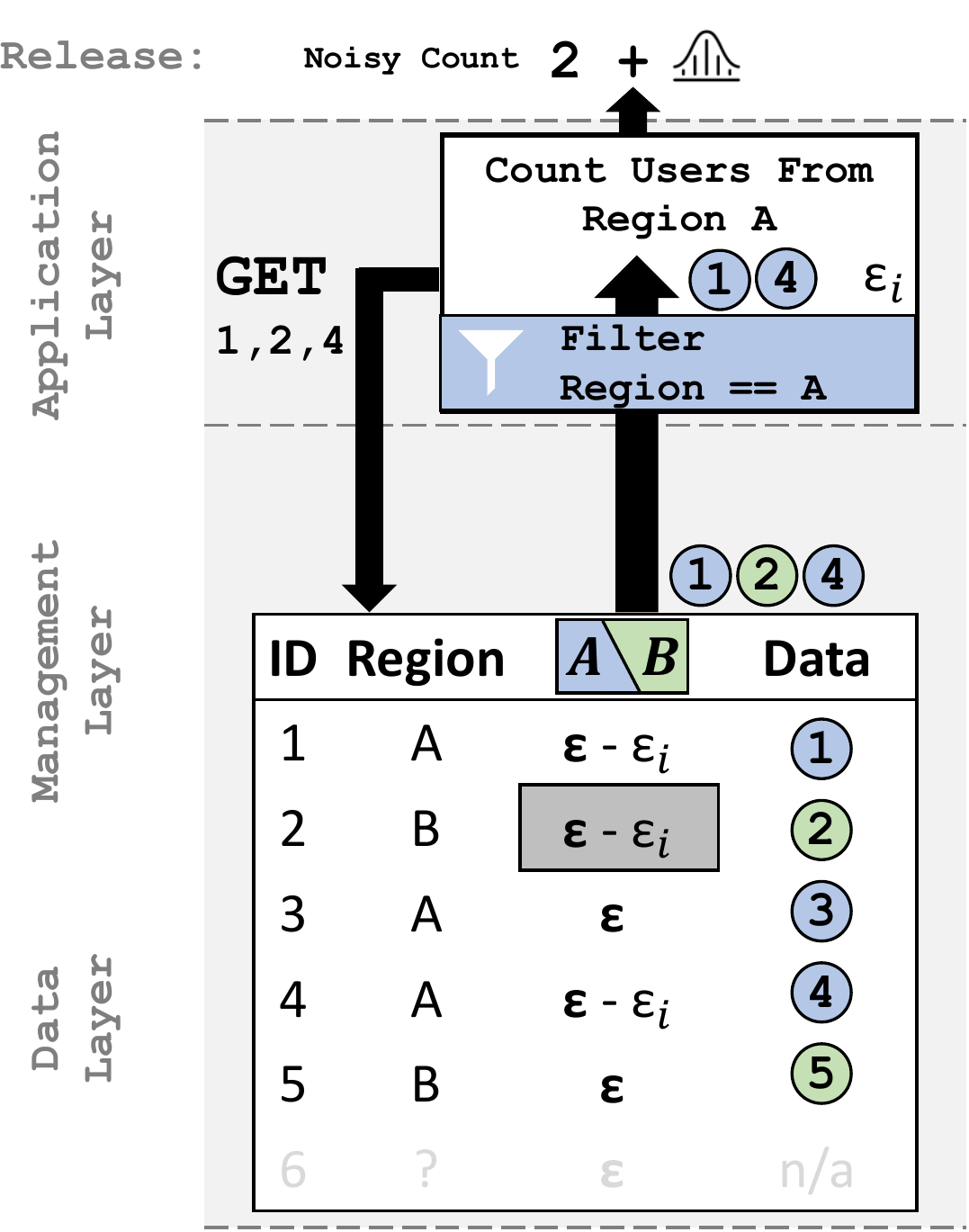}
        {\captionsetup{margin={0in,0.3in}}
        \caption{w/o PAs}
        \label{fig:filtering:nopa}}
    \end{subfigure}\hfill
    \begin{subfigure}[t]{.49\columnwidth}
        \includegraphics[height=5.1cm]{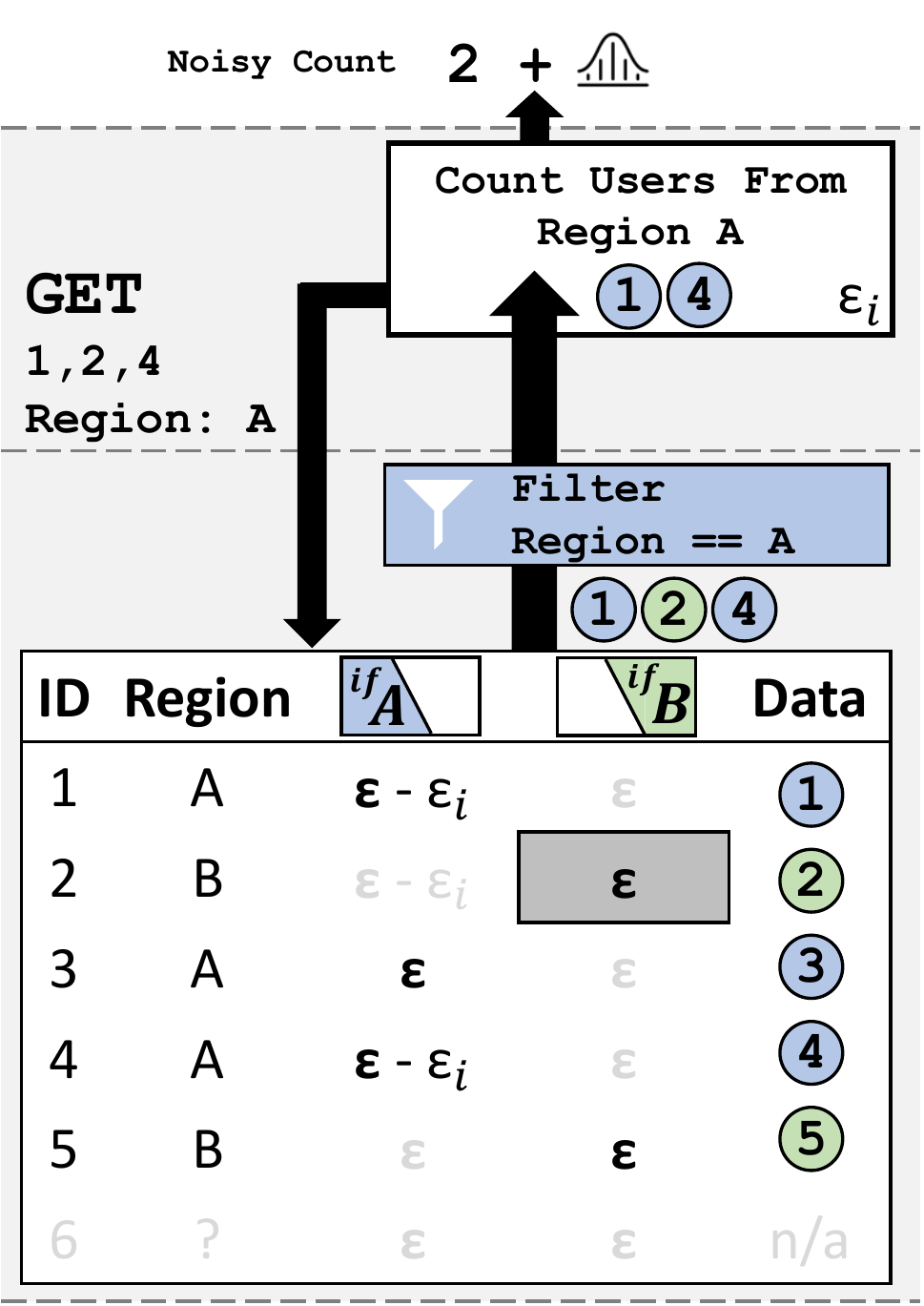}
        \caption{w/ PAs}
        \label{fig:filtering:pa}
    \end{subfigure}
    \caption{Using \glspl{pa} allows for more fine-grained budget accounting if the filtering happens in the management layer instead of the application.}
    \label{fig:filtering}
\end{figure}

The introduction of \glspl{pa} allows applications to express that they are only interested in users with certain attributes.
\Glspl{planner} without support for \glspl{pa} cannot utilize parallel composition and must deduct the privacy cost from all users assigned to the application (c.f. \Cref{fig:filtering:nopa}).
Here, the application would need to filter out users from Region B itself before computing and releasing the differentially private count.
However, a \gls{planner} with support for \glspl{pa} can integrate this information into the accounting process and apply the same filter before the data reaches the application layer (c.f. \Cref{fig:filtering:pa}), enabling more efficient accounting.
Note that the results of the applications will be identical between the two settings.
While \glspl{pa} improve the privacy analysis, they complicate the tracking of privacy budgets since the \gls{planner} needs to keep track of the privacy budget for each partition separately.
In the setting considered by Luo et al.~\cite{Luo2021-privacysched}, this is especially problematic since they already track a privacy budget per user and thus they would have to track a privacy budget for each user-partition pair.
However, the subsampling approach employed to improve the privacy analysis for applications operating on the same partition also simplifies tracking significantly.

\fakeparagraph{Subsampling}
Frequently, applications in large organizations do not need to run over the entire database but can already produce accurate results when run on a fraction of the users (e.g., many metrics).
Where this is the case, we can apply a technique known as \emph{amplification via subsampling}~\cite{Balle2018-dpsampling, Bun2022-dpsample} to reduce the privacy cost that is incurred.
Privacy amplification shows that it is both correct and generally more effective to account for the (reduced) cost of the query across the entire database rather than accounting for the exact cost of the query only on the selected users.
Specifically, given a population $\mathcal{U}$ of users in Poisson subsampling, we perform an independent Bernoulli trial for each user $i$ with probability $\gamma_i$ to include them in the sample.
The \gls{dp} mechanism is then applied to the sample, but the (significantly reduced) \gls{rdp} cost is accounted for across all users~\cite{Zhu2019-rdpsample}.
This also enables our system to omit the per-user tracking present in existing work~\cite{Luo2021-privacysched}.
In \Cref{sec:eval:sub}, we show the counter-intuitive benefits of subsampling, especially when considering many different such queries running over the same database.
DP applications could request more users and then apply subsampling internally, reporting the reduced privacy cost for the application.
Instead, \oursystem, integrates subsampling with \glspl{pa}, and allows requests to specify subsampling probabilities for partitions.
Integrating both partitioning attributes and subsampling within the privacy management layer enables a significantly tighter privacy analysis, as demonstrated in our evaluation (\cfref{sec:eval}).

\subsection{Privacy for Continuously Running Systems}
\label{sec:dp:cont}
\noindent
Even the most fine-grained privacy analysis will eventually deplete the available (finite) privacy budget in the context of a  system continuously applying DP mechanisms.
However, in large real-world systems, there is virtually always a significant influx of new users.
While privacy budgets for users must be finite (and comparatively small, in order to achieve strong guarantees), we can retire users and replace them with new users to replenish the overall available budget, as proposed in existing work~\cite{Luo2021-privacysched}.
However, user rotation introduces new challenges, especially if we want to both avoid the need to track individual user budgets and want to maximize how the budget can be used.

\fakeparagraph{User Rotation}
While there are many possible ways to implement user rotation, we need to consider the semantics (or lack thereof) that different approaches provide.
In addition, there are requirements for user rotation to maintain DP guarantees (c.f. \Cref{sec:priv-analysis}).
As a result, we consider three crucial properties: \emph{(i)} a user must be selected at most once, \emph{(ii)} the assignment of a user to an active set must be independent of other users, and \emph{(iii)} applications should have clear semantics for the active set that they are operating on.
Existing approaches suggest retiring users when their past privacy consumption reaches a certain threshold~\cite{Lecuyer2019-sage, Luo2021-privacysched}.
However, this method is problematic because it can lead to a biased population of active users without clear semantics, as the number of active users is dependent on prior queries.
The issue is amplified when paired with \glspl{pa}: if queries are mainly directed at specific subpopulations, then running queries across all users introduces a bias towards users who are queried less frequently since these users are the ones who remain.

In \oursystem, we instead propose a gradual and interpretable user rotation based on a \emph{sliding window} approach, where we consider the users to be partitioned into \glspl{group}, and in each round, retire the \gls{group} that was active the longest and activate a new one.
\!\footnote{Retired \glspl{group} with any remaining budget can be moved to a separate pool of resources available on a first-come, first-served basis by applications without the necessity of clear semantics.}
This approach removes the dependency on prior queries and thus eliminates the potential bias.
Given $K$ active groups, the budget in each round should be roughly $\frac{1}{K}$ of the overall budget.
One trivial instantiation would be to group users by the round during which they joined, however, limiting queries to only data from the last $K$ round is too restrictive for many scenarios.
Instead, we can randomly assign users that join during the current round to groups corresponding to the next $T$ rounds (for $T\gg K$).
With this approach, a \gls{group} is always an unbiased sample of the user base joining in the last $T$ rounds.
For example, with weekly allocations, $T=104$ and $K=12$, we always have data spanning the last two years, with (on average) as many users as the system sees joining in a quarter.
In essence, \oursystem provides sliding window semantics for users across all applications.
While a certain distribution shift is inevitable when retiring and introducing new users, our careful user-to-group assignment prevents sudden distribution shifts.
This represents a significant improvement over the existing state-of-the-art that activates users in arrival order, potentially leading to an arbitrarily biased active user set.
For instance, consider a marketing campaign targeting a specific country, resulting in a temporary sudden surge of users from that region.
In \oursystem, these users would be distributed evenly across the next $T$ groups, rather than being disproportionately concentrated in a single group.

\fakeparagraph{Budget Unlocking}
Whenever a new group is added to the active set, the budget for these users becomes available and should then be consumed roughly evenly across the next $K$ rounds.
Specifically, we always want the oldest (technically, the most depleted) group in the active set to have roughly $\frac{1}{K}$ budget left, so that we can guarantee a certain level of budget is always available for requests in each round.
To provide such guarantees, we need to keep part of the budget locked for future rounds.
If $\epsilon$ is the total budget, then the obvious strategy is to unlock $\frac{\epsilon}{K}$ per round. \!\footnote{We illustrate budget unlocking with a scalar budget $\epsilon$, but in reality, we have a vector of budgets due to \gls{rdp} accounting.}
Since in every round additional budget is unlocked for all active \glspl{group}, we are guaranteed to have at least $\frac{\epsilon}{K}$ of the budget available, satisfying our balancing concerns.
In the context of Poisson subsampling, we will in fact always have \emph{exactly} this much budget available, as subsampling is limited by the lowest available budget in the active set\footnote{We could apply non-uniform subsampling with different sampling probabilities across \glspl{group}, however, this could result in unclear semantics.}.
As a result, any unused budget from one round cannot be used in later rounds as that would cause a round to consume more than $\frac{\epsilon}{K}$.

We address this issue by introducing a novel budget-unlocking strategy, which biases the unlocking of the budget towards the first half of the rounds a group remains active.
Essentially, we can use the budget from future rounds ``on credit''.
Clearly, the bias should be subtle to provide fairness across rounds, but even a subtle bias is sufficient to allow leftover budget from older groups to be consumed (\cfref{sec:eval:unlock}).
In the first half of the rounds, we unlock $(1 + \Delta) \frac{\epsilon}{K}$ in every round, and in the second half of the rounds, we unlock $(1 - \Delta) \frac{\epsilon}{K}$ in every round where $\Delta \in [0, 1]$.
More formally the unlocked budget in round $1 \leq k \leq K$ is then given by:

\begin{equation}
    \label{eq:unlock}
   \epsilon_k =  \frac{\epsilon}{K}  \cdot  \left( k + \sum_{i=1}^{\min(k, \lfloor \frac{K}{2} \rfloor)} \Delta - \sum_{i=\lceil \frac{K}{2} \rceil + 1}^{k} \Delta \right)
\end{equation}

At $k=K$, the budget is fully unlocked, i.e., $\epsilon$.

\begin{theorem}
    The available budget in any round is always in between $(1 - \Delta) \frac{\epsilon}{K}$ and $(1 + \Delta) \frac{\epsilon}{K}$.
\end{theorem}
\noindent
The proof is straightforward.
The most recent active \gls{group} has an unlocked budget of $(1 + \Delta) \frac{\epsilon}{K}$.
Since any allocation always needs to satisfy the constraints of all \glspl{group}, no more than $(1 + \Delta) \frac{\epsilon}{K}$ budget can be consumed in a round.
For any active \gls{group} we unlock at least $(1 - \Delta) \frac{\epsilon}{K}$ additional budget and thus we have at least $(1 - \Delta) \frac{\epsilon}{K}$ budget available in every round.
In \Cref{sec:appendix:unlock}, we provide more detailed proofs of our approach's properties.

\fakeparagraph{Setting Hyperparameters}
\oursystem introduces several hyperparameters that warrant careful consideration in practical deployments.
Of these, the global target $(\epsilon, \delta)$ privacy guarantee is certainly the most important hyperparameter.
Yet, the strategies for user rotation and budget unlocking also introduce new hyperparameters, which may seem challenging to configure.
To aid in this process, we provide a tool that enables practitioners to specify their requirements - for instance, the expected number of active users needed in each round - and visualize the effects of different hyperparameter combinations on the system.
Additionally, practitioners can utilize our workload generator (\cfref{sec:workload}) to explore combinations of mechanisms permitted within a round.

\subsection{Privacy Request Allocation}
\label{sec:alloc}
\noindent
\Gls{dp} naturally limits the number of requests that can be executed on the data of a given active set of users.
Due to the complex and non-linear nature of DP composition, especially in the context of partitioning attributes, it is a non-trivial optimization problem to determine which combination of requests will make the best use of the budget.
In fact, we would not want to optimize for the fullest use of the budget, but rather for a more meaningful metric.
While maximizing the \emph{number of allocated requests} might seem natural, the results of DP applications frequently improve exponentially with linear increases in budget, making this metric undesirable.
Instead, we aim to maximize the total \emph{utility} of the accepted requests, which also provides a way to express priorities between different applications which might be critical for real-world deployments.

\fakeparagraph{Optimization Problem}
Choosing a set of requests that maximizes utility under a budget constraint resembles a type of Knapsack problem.
The non-linear composition of privacy budgets, especially with partitioning attributes, requires us to consider a \emph{multidimensional} Knapsack, where each partition is its own dimension.
In addition, RDP-composition tracks many different variants of the DP bounds (different $\alpha$-orders) for each partition and it is sufficient for only one to be satisfied.
Tholoniat et al. present an efficient heuristic for this optimization problem~\cite{Tholoniat2022-dppacking}.
However, naively incorporating \glspl{pa} along with subsampling, user rotation, and budget unlocking would result in a significantly higher-dimensional problem instance compared to prior work with a simpler setting~\cite{Luo2021-privacysched, Tholoniat2022-dppacking}.
We counter-act this expansion with optimizations that allow us to collapse many of the newly introduced dimensions, e.g., by considering only the \glspl{pa} present in the current set of requests.
We first present a formal definition of the optimization problem in \oursystem and then discuss our optimizations.

Let $\mathcal{R} \coloneqq \{1, \; \ldots, \; R_{max}\}$ be the index set of requests.
A request $R_i$ for $i \in \mathcal{R}$ is a tuple $(\Phi_i,\mathbf{C}_i,W_i)$ consisting of a \emph{data requirement} defining a subpopulation encoded as a propositional formula $\Phi_i$ over the \glspl{pa},
a \emph{budget requirement} expressed as a vector $\mathbf{C}_i \in \mathbb{R}_{\geq 0}^{|\mathcal{A}|}$ of \gls{rdp} costs for different orders $\alpha_a$ (for $a \in \mathcal{A} \coloneqq \{1, \; \ldots, \; A_{max}\}$),
and a \emph{weight} $W_i \in \mathbb{N}$.
We introduce decision variables $y_i \in \{0, 1\}$ for $i \in \mathcal{R}$, where $y_i = 1$ means the request $R_i$ is accepted, and $y_i = 0$ means the request has been rejected.
Let $\mathcal{S} \coloneqq \{1, \ldots, S_{max}\}$  denote the index set over all \emph{blocks}.
A block $S_j$ for $j \in \mathcal{S}$ is a tuple $(\gls{groupid}, \Psi_j, \mathbf{B}_j)$ consisting of a \gls{groupid}, a propositional formula $\Psi_j$ over the \glspl{pa} and a remaining budget $\mathbf{B}_j \in \mathbb{R}_{\geq 0}^A$ for this block in this \gls{group}.
The remaining budget $\mathbf{B}_j$ of a block is determined by the difference of the unlocked budget according to \Cref{eq:unlock}, and the accumulated allocation history.

We set the demand for a block $d_{ij}^{(\alpha)} = \mathbf{C}_i^{(\alpha)}$ if $\Phi_i \land \Psi_j$ is satisfiable, and $0$ otherwise.
This captures the fact that the demand for a block outside the selected subpopulation of a request is 0.
Finally, to find the optimal set of requests to accept, we solve the optimization problem:

\begin{align*}
        & \underset{}{\text{max}} &  &  \smashoperator{\sum_{i \in \mathcal{R}}} \ y_i \cdot W_i  &  & \\
        & \text{s.t.}  & & \sum_{i \in \mathcal{R}}  d_{ij}^{\;(\alpha)} y_i \leq B_j^{\;(\alpha)}   && [\forall j \in \mathcal{S} \;\; \exists \alpha \in \mathcal{A}] \\
\end{align*}

\noindent
We also formulate a more general optimization problem (as an \gls{ilp}) that supports alternative allocation approaches in \Cref{sec:appendix:ilp}.
Practical instances of this optimization problem are within the scope of commercial \gls{ilp} solvers~\cite{Gurobi2022-doc}.
Alternatively, Tholoniat et al. propose a heuristic for the knapsack problem variant which arises in \gls{rdp} block composition~\cite{Tholoniat2022-dppacking}.
However, we present optimizations that reduce the dimensionality of the problem which makes solving it significantly more efficient.

\begin{figure}[t]
    \centering
    \includegraphics[width=0.85\columnwidth]{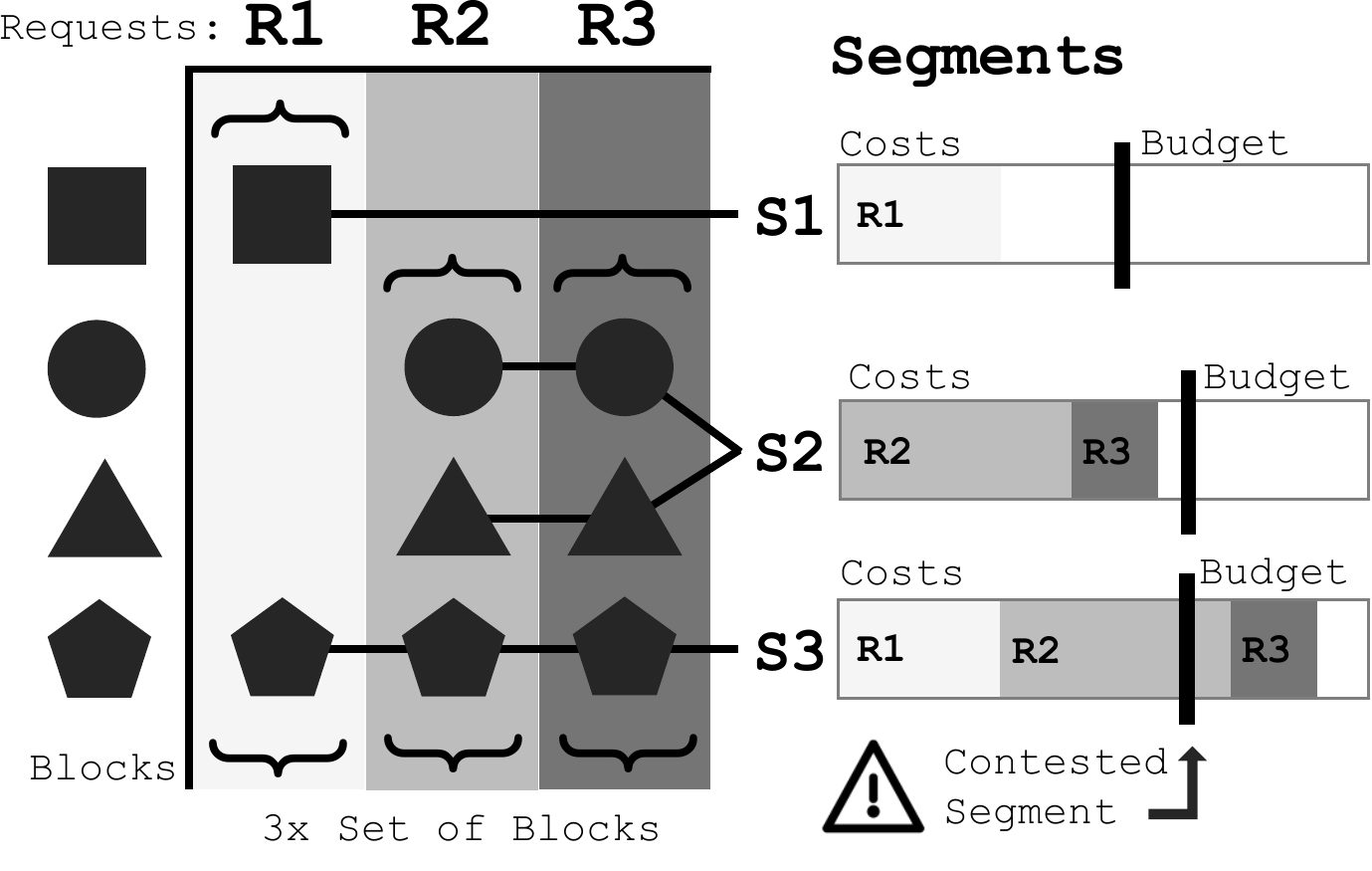}
    \caption{
        An example of three requests R1, R2, and R3 selecting four blocks.
        Vertically, we show the set of blocks selected by a request.
        Horizontally, we show blocks grouped into segments defined by a unique set of interested requests.
    }
    \label{fig:segments}
\end{figure}

\fakeparagraph{Solving the Scalability Issues}
The number of dimensions in the optimization problem as described above is equal to $K \cdot \left( \left|\texttt{A}_\texttt{1}\right| \times \cdots \times \left|\texttt{A}_\texttt{k}\right| \right)$, where $\left|\texttt{A}_\texttt{i}\right|$ is the domain size of the $i$th PA.
As a result, this can quickly explode for fine-grained partitioning attributes.
We introduce two optimizations to reduce the problem size:
\emph{(i)} we consolidate \glspl{blockid} into \emph{segments} based on the current batch of requests,
and \emph{(ii)} we identify segments and requests that can safely be excluded from the optimization problem.
\oursystem computes \emph{segments} that succinctly represent groups of \glspl{blockid} relevant to the current set of candidate requests, drastically reducing the number of privacy budgets we need to consider.

Our solution exploits that the number of unique combinations of the \glspl{pa} in the current requests is significantly smaller than the number of total possible combinations.
A request's selection of \glspl{pa} directly defines a set of `compatible' blocks (shown vertically in \Cref{fig:segments}).
\emph{Segments} are sets of blocks which (for the current set of requests) will be treated identically by any allocation.
For example, in \Cref{fig:segments}, where segments are illustrated using horizontal lines, we see that if the \mycircle{black} block is allocated to a given request, so will the \mytriangle{black} block.
Therefore, those two blocks form a segment (S2).
Since the blocks in a segment behave identically under allocation, \oursystem's \gls{planner} only requires the per-segment privacy budget for each \gls{group}, significantly reducing the constraints it has to consider.

Additionally, we reduce the number of segments that we need to consider by focusing only on \emph{contested} segments.
A segment is contested if accepting all requests associated with it would exceed the privacy budget of the segment (e.g., S3 in \Cref{fig:segments}).
We ignore non-contested segments in our problem formulation since their privacy budgets will be satisfied no matter the chosen allocation.
Further, we always accept requests that are associated with only uncontested segments.
Conversely, we always reject requests that by themselves violate the privacy budget of any associated segment.
Therefore, we eliminate both groups of requests.

\subsection{Privacy Analysis}
\label{sec:priv-analysis}
\noindent
\oursystem uses block composition~\cite{Lecuyer2019-sage} and instantiates it with an \gls{rdp} privacy filter for each block.
Under block composition, the overall mechanism satisfies $(\epsilon, \delta)$-\gls{dp}, as long as the $(\epsilon, \delta)$-\gls{dp} budget is not exceeded on any block.
In \oursystem, each block, i.e., partition, corresponds to a \gls{blockid} and there is a block for every possible combination of \glspl{pa} and for each \gls{group}.
The assignment from users into \glspl{group} (\cfref{sec:dp:cont}) and to a block within that group based on the users' \glspl{pa} is immutable and every user is only assigned to one block.
The \glspl{pa} (i.e., years, regions) have a public domain, and the assignment from users to \glspl{group} is only based on their arrival time, and importantly, independent of other users' assignments.
As a result, every user only influences data releases computed on that user's block, satisfying the requirements of block composition~\cite{Lecuyer2019-sage}.

\oursystem's \gls{planner} is data-independent, meaning it never touches sensitive data.
More concretely, the inputs are always public knowledge (schema of \glspl{pa}, request batch, and the currently available \glspl{group}, and their allocation history).
Consequently, the \gls{planner} always sees the same number of blocks in each round.
Considering that future mechanisms and budget choices can depend on previous results, our privacy analysis for the data releases must consider the fully adaptive setting (\cfref{sec:bg:dp:comp}).
Moreover, as we allocate budgets to various mechanisms without enforcing a sequential consumption of these budgets, our privacy analysis is in the concurrent composition setting~\cite{Vadhan2022-dpconcurrent}.
The \gls{rdp}-based filter we employ meets the requirements for fully adaptive concurrent composition~\cite{Lecuyer2021-dpodometers, Haney2023-dpfadaptconc}.

\newglossaryentry{mouse}{name={mouse}, plural={mice}, description={}}
\newglossaryentry{hare}{name={hare}, plural={hares}, description={}}
\newglossaryentry{elephant}{name={elephant}, plural={elephants}, description={}}

\newacronym{gm}{\texttt{GM}}{Gaussian Mechanism~\cite{Dwork2006-dpgaussian}}
\newacronym{lm}{\texttt{LM}}{Laplace Mechanism~\cite{Dwork2006-originaldp}}
\newacronym{svt}{\texttt{SVT}}{Sparse Vector Technique~\cite{Dwork2009-svt}}
\newacronym{rr}{\texttt{RR}}{Randomized Response~\cite{Warner1965-randresp}}
\newacronym{nsgd}{\texttt{N-SGD}}{NoisySGD Mechanism~\cite{Abadi2016-mldp}}
\newacronym{pate}{\texttt{PATE}}{PATE~\cite{Papernot2018-scalablepate}}
\newacronym{zcdp}{\texttt{zCDP}}{zCDP Mechanism~\cite{Bun2016-zCDP}}

\newacronym{upc}{UPC}{User-Parallel Composition}

\subsecspacingtop
\section{Evaluation}
\subsecspacingbot
\label{sec:eval}

\label{sec:eval:sub}
\label{sec:eval:unlock}

\noindent
In this section, we evaluate the effectiveness of \oursystem's fine-grained privacy analysis across different workloads. We focus our evaluation on the planning phase (i.e., the \gls{planner}), which is the core component of \oursystem's design and effectively handles most of the complexity of DP management. Note that other components in our system correspond to widely deployed existing tools (e.g., Lakehouse, Data tagging, ABAC enforcement).

\subsection{Evaluation Setup}
\label{sec:workload}

\fakeparagraph{DP Workload Generator}
Benchmarking \gls{dp} systems remains a major challenge~\cite{Cummings2023-dpfrontier} and this difficulty is exacerbated when dealing with mixed workloads.
Our work tackles this challenge by introducing a configurable DP workload generator that can accommodate various request types and workload characteristics.
The key to its flexibility lies in recognizing that any DP application can be represented as a combination of a small number of fundamental DP mechanisms.
For example, training an ML model with DP-SGD~\cite{Abadi2016-mldp} can be represented as a combination of Gaussian mechanisms.
By doing so, we are able to abstract a diverse range of real-world use cases into a more generalized setting.
We make our workload generator available as an open-source tool, as we believe better benchmarks for complex mixed DP workloads are of independent interest.

Our workload generator offers parameters that enable users to fine-tune workload characteristics, tailoring them to specific deployment scenarios.
These parameters offer control over aspects such as the frequency of allocation rounds, arrival rates of users and requests, and the schema of \glspl{pa}.
At the core of our generator is the request distribution, which models diverse request families as a categorical distribution over request types.
Request types are parameterized with data and privacy requirements, allowing for granular customization and fine-grained management of workload dynamics.
Data requirements are sampled from a categorical distribution over partitioning attributes, defining a target population, and a categorical distribution over the percentage of that population to be requested.
For example, a request type might require either 10\% or 50\% of all active users above age 50 with equal probability.
Privacy requirements specify a DP mechanism (e.g., Laplace Mechanism) with privacy costs sampled from a categorical distribution of possible costs.
For example, a request type might specify the DP-SGD mechanism and an equal split between low and high privacy costs, modeling, e.g., different numbers of epochs.
DP mechanisms are represented using the Auto DP library~\cite{Wang2021-autodp}.
Using a discrete event simulator~\cite{simpy}, we generate random workload instances for a specific deployment scenario.
Finally, we provide adapters that transform samples into concrete requests for specific budget management systems (e.g., \oursystem or PrivateKube~\cite{Luo2021-privacysched}).

\fakeparagraph{Instantiating the Utility Distribution}
We assign each request a scalar \emph{utility} to represent its expected utility to the organization.
We model their distribution as a Cobb-Douglas production function~\cite{Cobb1928-cobbdouglas} $Y=A \cdot L^{\beta }K^{\alpha}$, which is traditionally used to model production output as a function of labor (L) and capital (K) at a given total factor of productivity (A).
We instantiate L with the privacy cost and K with the amount of data requested. Then, the function correctly predicts that with no budget or no data, no utility can be generated and that utility increases as either one or both increase.
In the function, $\alpha$ and $\beta$ describe the elasticity of the relation, and we set $\alpha = 1$ and $\beta = 2$, which roughly models that increase in privacy budget are usually more impactful than increases in the amount of data.\!\footnote{
    A linear change in $\epsilon$ translates to an exponential increase in the range of allowed distributions.
}
For the privacy cost, we employ a mechanism-dependent correction factor in the utility assignment applied to each request's privacy cost, such that requests from the same cost category also share the same privacy cost to calculate utility.
We vary the productivity~$A$ randomly according to the following distribution $A \sim \text{Beta}(0.25, 0.25)$ to model application-level variations in utility.
For example, recommender systems can be useful even when relatively inaccurate, while fraud detection systems must achieve higher accuracy to provide value.
Finally, we normalize the sum of the utility of each workload to ensure comparability between workloads.

\fakeparagraph{Evaluation Scenario}
For our evaluation, we instantiate the workload generator in a setting modeling weekly allocations.
We simulate a total of 40 weeks using a sliding window of 12 rounds (i.e., weeks) to demonstrate the continuous operation of \oursystem.
Request arrivals are modeled as a Poisson process with a mean inter-arrival time of 20 minutes, resulting in an expected 504 requests per allocation round.
We also model user arrivals as a Poisson process (inter-arrival time 10s), however, we note that the frequency and absolute number of users joining the system does not affect the number of \glspl{group} that need to be tracked~(\cfref{sec:dp:cont}).

For data requirements, we instantiate the \gls{pa} domain as a vector of size 204'800.\footnote{Note any number of \glspl{pa} can be flattened to a single but larger \gls{pa}.} %
We generate the \gls{pa} selection for a request as a consecutive range of $\lfloor s \cdot 100 \rfloor$ elements in the \gls{pa} vector, at a location chosen uniformly at random and wrapping around the edges, with  $s \sim Beta(a, b)$.
Since real-world workloads do not all partition equally well in practice, we explore different distributions for selecting the requested population based on \glspl{pa}.
The parameters $a$ and $b$ allow precise control over the distribution of \gls{pa} overlaps.
Requests require either 25\%  or 100\% of the selected population with equal probability.

We enforce a global user-level \gls{dp} guarantee with a budget of $\epsilon = 3$ and $\delta = 10^{-7}$.
and we instantiate the \gls{rdp} privacy filter with the following orders $\{1.5, 1.75, 2, 2.5, 3, 4, 5, 6, 8, 16, 32, 64, 10^6, 10^{10}\}$.
We construct our workloads as combinations of the \gls{gm}, \gls{lm}, \gls{svt}, \gls{rr}, \gls{nsgd}, and \gls{pate}, capturing common mechanisms from a wide range of domains.
We group requests into three categories occurring with equal probability: \glspl{mouse}, \glspl{hare}, and \glspl{elephant}, representing low, medium, and high privacy costs, respectively.
For \gls{gm}, \gls{nsgd}, and \gls{pate}, we use $\epsilon = 0.05, 0.2, 0.75$ for \glspl{mouse}, \glspl{hare}, and \glspl{elephant}, respectively.
While for \gls{lm}, \gls{svt}, and \gls{rr}, we use $\epsilon = 0.01, 0.1, 0.25$ to compensate for the less efficient accounting under \gls{rdp}.
We keep $\delta$ fixed at $10^{-9}$ for all mechanisms.
Given an $(\epsilon, \delta)$ target, we derive a corresponding \gls{rdp} vector for each mechanism~\cite{Wang2021-autodp}.

\fakeparagraph{Workloads}
We evaluate four workloads combining different data and privacy requirements modeling a range of real-world scenarios in which \oursystem might be deployed.
\newglossaryentry{w:count}{name={W1:GM}, description={
            The first workload consists purely of requests using the Gaussian Mechanism (GM) representing, e.g., the predicate counting queries over the Census dataset~\cite{Sexton2017-censussyn, Kotsogiannis2019-dpsql}.
            These queries are of the form SELECT Count(*) FROM x WHERE $\Phi$, where $\Phi$ is a predicate over the \glspl{pa}.
            This is a best-case setting, as the GM is especially amenable to \gls{rdp} composition.
            Further, this workload is heavily partitioned, i.e., $s \sim Beta(1, 10)$, and can, therefore fully use parallel composition.
        }}

\newglossaryentry{w:fund}{name={W2:Mix}, description={
            We consider a more varied mix of mechanisms in the second workload.
            Specifically, the workload uses an equal distribution of \gls{gm}, \gls{lm}, \gls{svt}, and \gls{rr} mechanisms, representing important key building blocks for instance, used in database query tasks.
            For \gls{gm} and \gls{lm}, we consider the heavily partitioned predicate counting queries from the previous workload.
            For \gls{svt} and \gls{rr}, we consider requests that cannot be easily partitioned (i.e., $s \sim Beta(1,0.5)$) as this naturally matches these systems.
            This workload represents a more realistic challenge as these mechanisms do not always compose efficiently, and half of the requests barely exploit \glspl{pa}.
        }}

\newglossaryentry{w:ml}{name={W3:ML}, description={
            The third workload features a machine learning workload consisting of an equal mix between \gls{nsgd} and \gls{pate} based ML training.
            In this workload, a typical query selects half the domain, i.e., $s \sim Beta(2, 2)$, and thus generates a significant overlap between requests.
            This models the fact that ML tasks usually do not have as many natural restrictions on the data they consider and tend to require large amounts of data in order to achieve acceptable results.
            We thus expect that, despite the same privacy costs, fewer of the requests in this workload can be allocated.
        }}

\newglossaryentry{w:all}{name={W4:All}, description={
For the last workload, we consider an equal combination of requests from all other workloads.
Each mechanism is instantiated using the same parameters as described in the respective workloads.
This workload models the wide variety of different types of \gls{dp} applications that we would expect in a practical deployment in a large organization.
}}

\noindent
(\underline{\Gls{w:count}}): \glsdesc{w:count}

\noindent
(\underline{\Gls{w:fund}}): \glsdesc{w:fund}

\noindent
(\underline{\Gls{w:ml}}): \glsdesc{w:ml}

\noindent
(\underline{\Gls{w:all}}): \glsdesc{w:all}

\noindent

\begin{figure}[t]
    \vspace{3mm}
    \begin{subfigure}[b]{0.48\columnwidth}
        \centering
        \includegraphics[trim={0 22pt 0 0}, clip, width=1.0\linewidth]{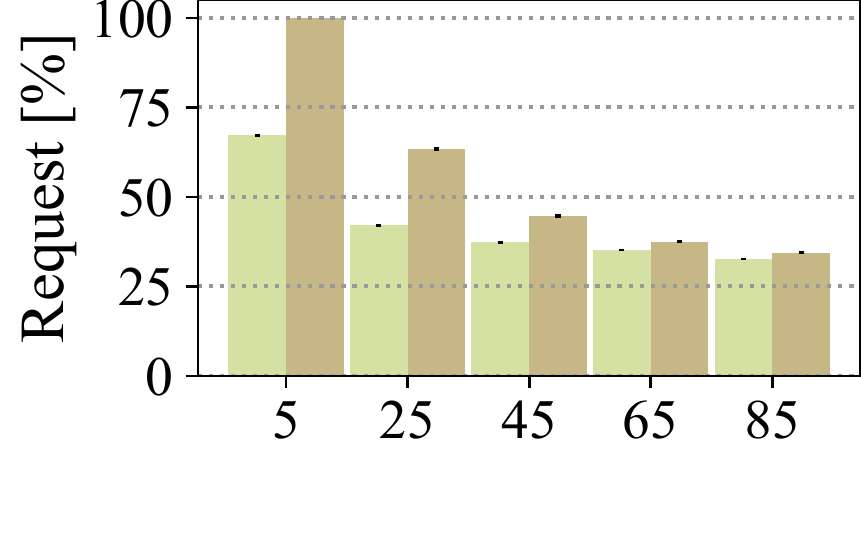}
        \includegraphics[width=1.0\linewidth]{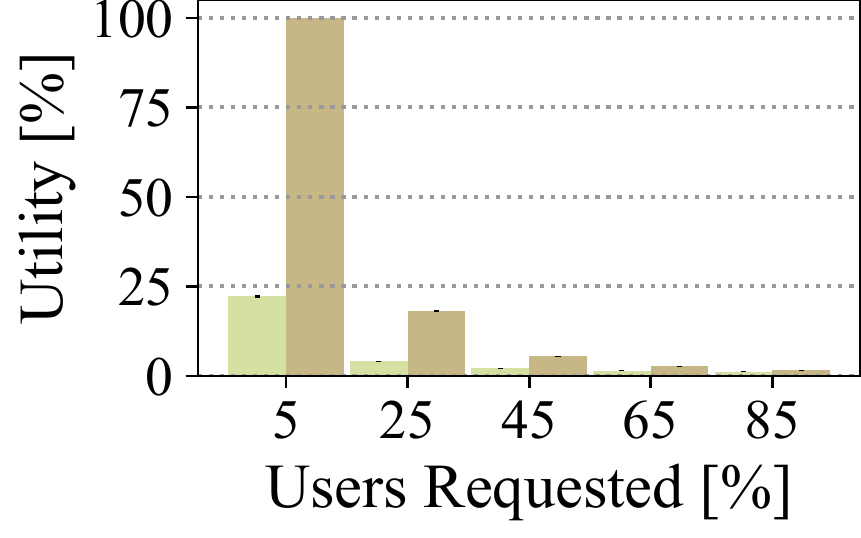}
        \caption{\gls{w:count}}
    \end{subfigure}
    \begin{subfigure}[b]{0.48\columnwidth}
        \includegraphics[trim={20pt 22pt 0 0}, clip, width=\dimexpr\linewidth-9pt\relax]{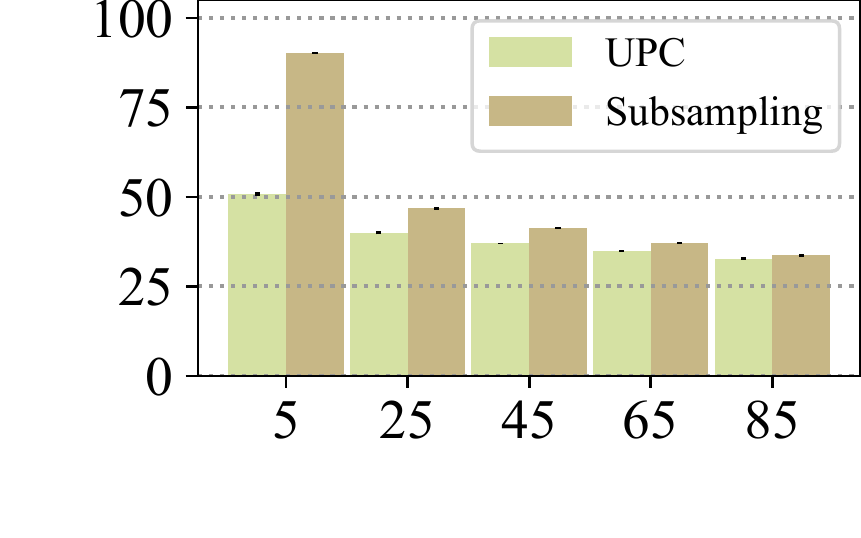}
        \includegraphics[trim={20pt 0 0 0}, clip, width=\dimexpr\linewidth-9pt\relax]{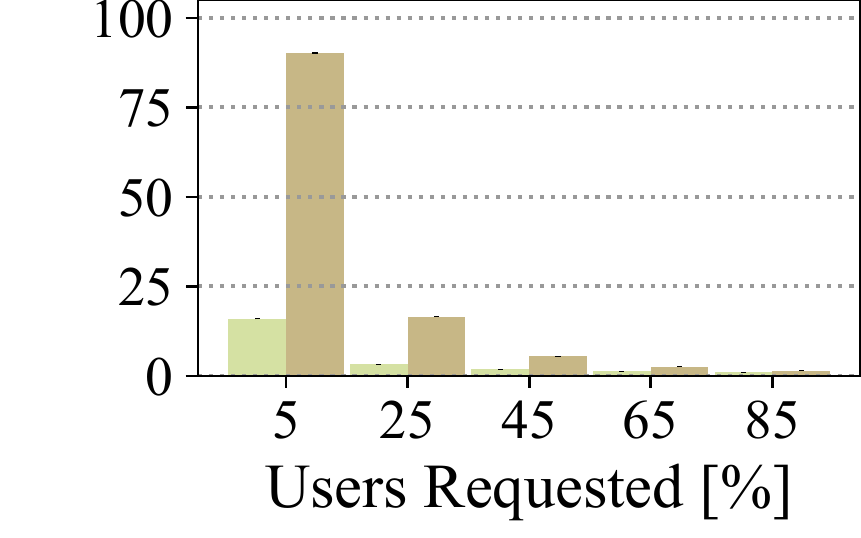}
        \caption{\gls{w:fund} }
    \end{subfigure}
    \caption{Comparing \acrshort{upc} with Poisson subsampling on two workloads, varying the fraction of the selected population.}
    \label{fig:sub}
\end{figure}

\begin{figure}[t]
    \centering
     \begin{subfigure}[b]{1.0\columnwidth}
        \includegraphics[trim={0 20pt 0 0}, clip, width=1.0\columnwidth]{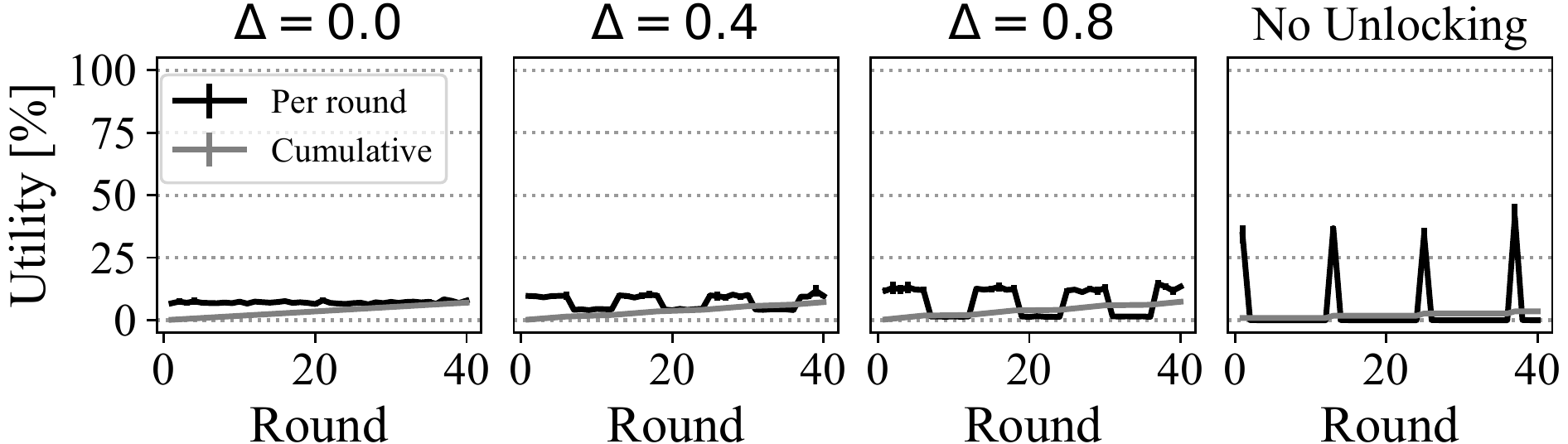}

    \caption{w/o PA}
        \vspace{-3pt} %
    \end{subfigure}

    \vspace{10pt}
    \begin{subfigure}[b]{1.0\columnwidth}
        \includegraphics[trim={0 0 0 18pt}, clip, width=1.0\columnwidth]{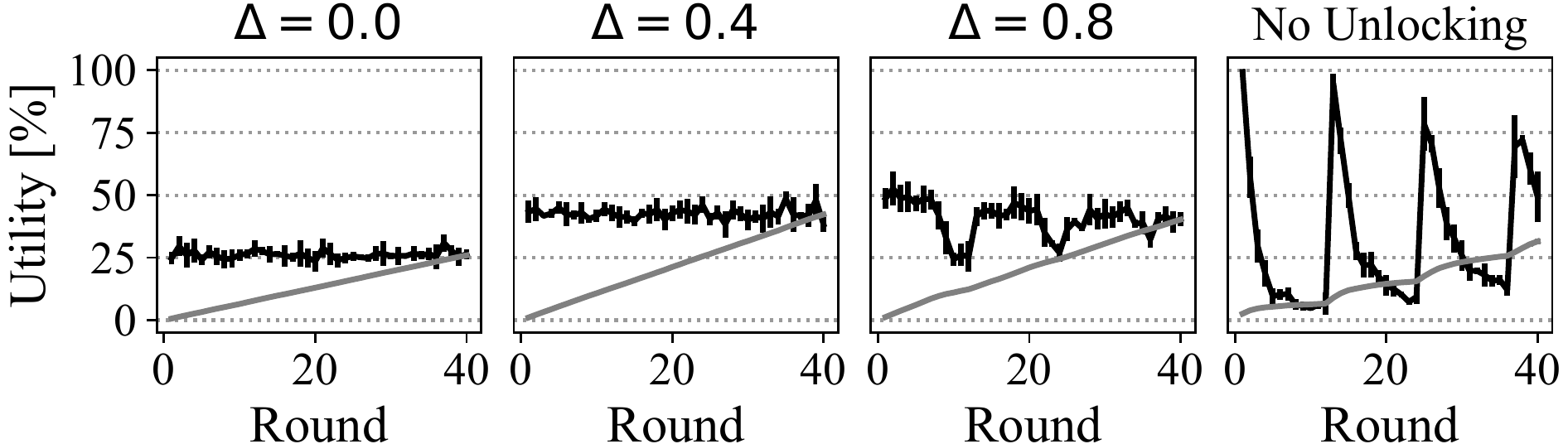}

            \caption{w/ PA}
        \end{subfigure}

    \caption{The effect of budget unlocking on utility per round achieved by the \gls{dpk} algorithm on the \gls{gm} workload.}
    \label{fig:unlock}
\end{figure}

\fakeparagraph{Implementation \& Experimental Setup}
We implement a prototype of \oursystem in Rust, consisting of roughly 13k SLOC.
\!\footnote{\oursystem's code available at: \url{https://github.com/pps-lab/cohere}}
Given the set of candidate requests of an allocation round (generated by the workload generator), the \gls{planner} computes the segments~(\cfref{sec:dp:pa}), formulates the optimization problem~(\cfref{sec:alloc}), and uses an allocation algorithm to accept requests.
Where we solve an ILP optimization problem (\cfref{sec:appendix:ilp}), we employ the Gurobi solver~\cite{Gurobi2022-doc}.

We run \oursystem on 16 virtual cores with 16~GB of memory on a scientific cluster backed by host machines with 128-core AMD EPYC CPUs (2.60GHz), with 256 GB memory running CentOS~7.
We evaluate the performance of the allocation algorithms on two main objectives: maximizing the percentage of requests and maximizing the percentage of total utility.
We perform the allocation for each workload on five instances randomly sampled by the workload generator and
report the mean and standard deviation across the five instances of each workload.

\begin{figure*}[t]
    \begin{subfigure}[t]{0.24\textwidth}
        \includegraphics[width=1.1\linewidth]{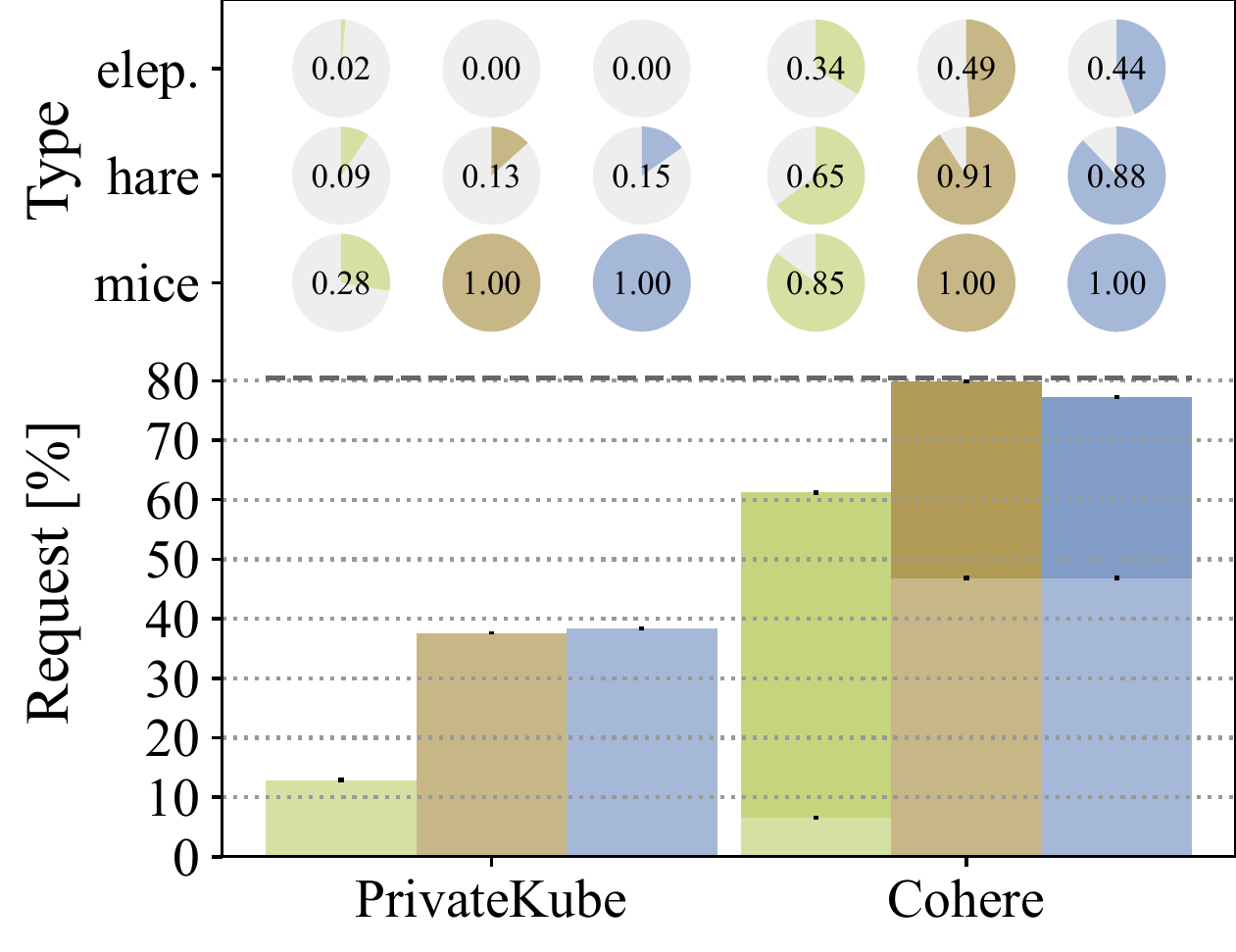}
    \end{subfigure}
    \begin{subfigure}[t]{0.24\textwidth}
        \includegraphics[width=1.1\linewidth]{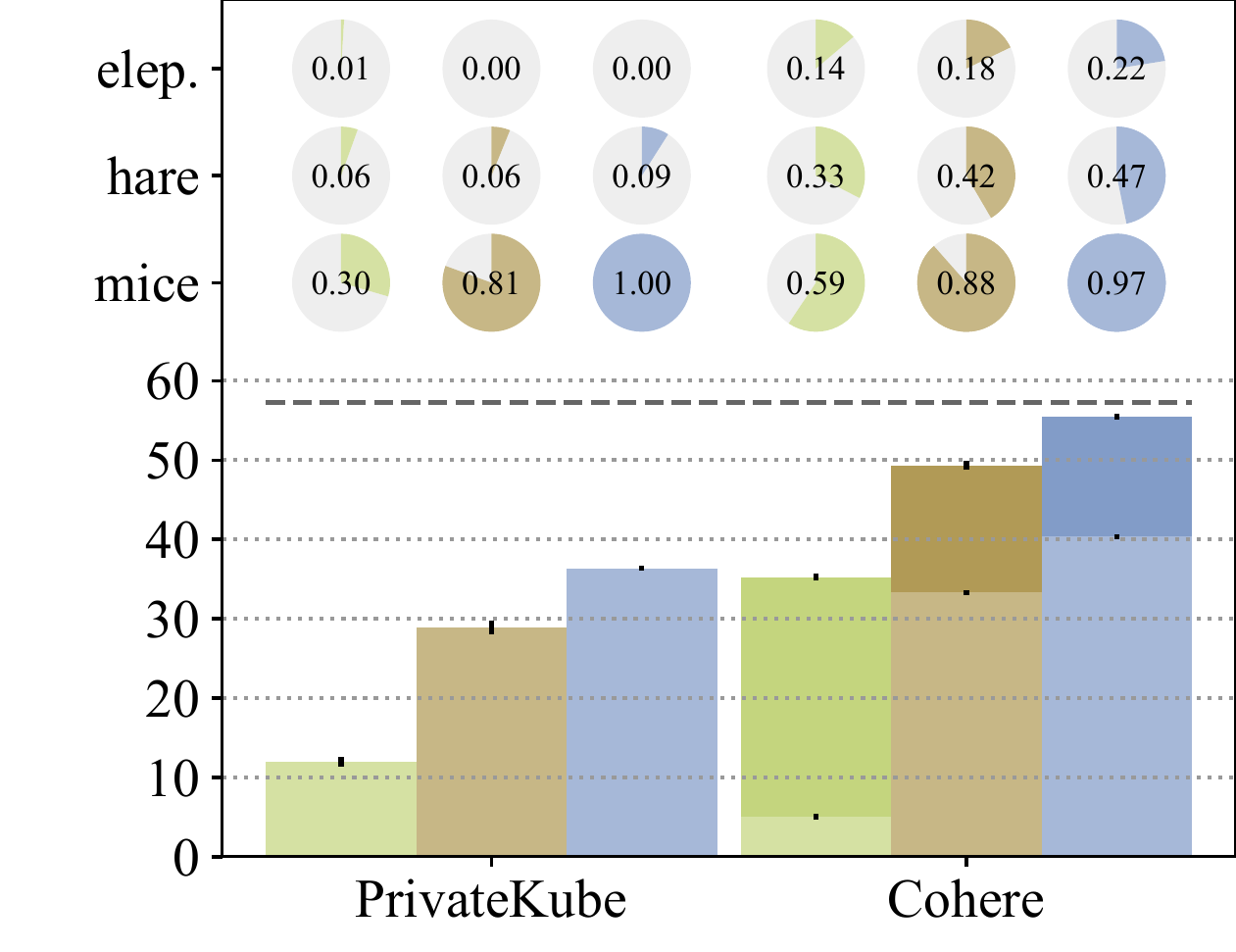}
    \end{subfigure}
    \begin{subfigure}[t]{0.24\textwidth}
        \includegraphics[width=1.1\linewidth]{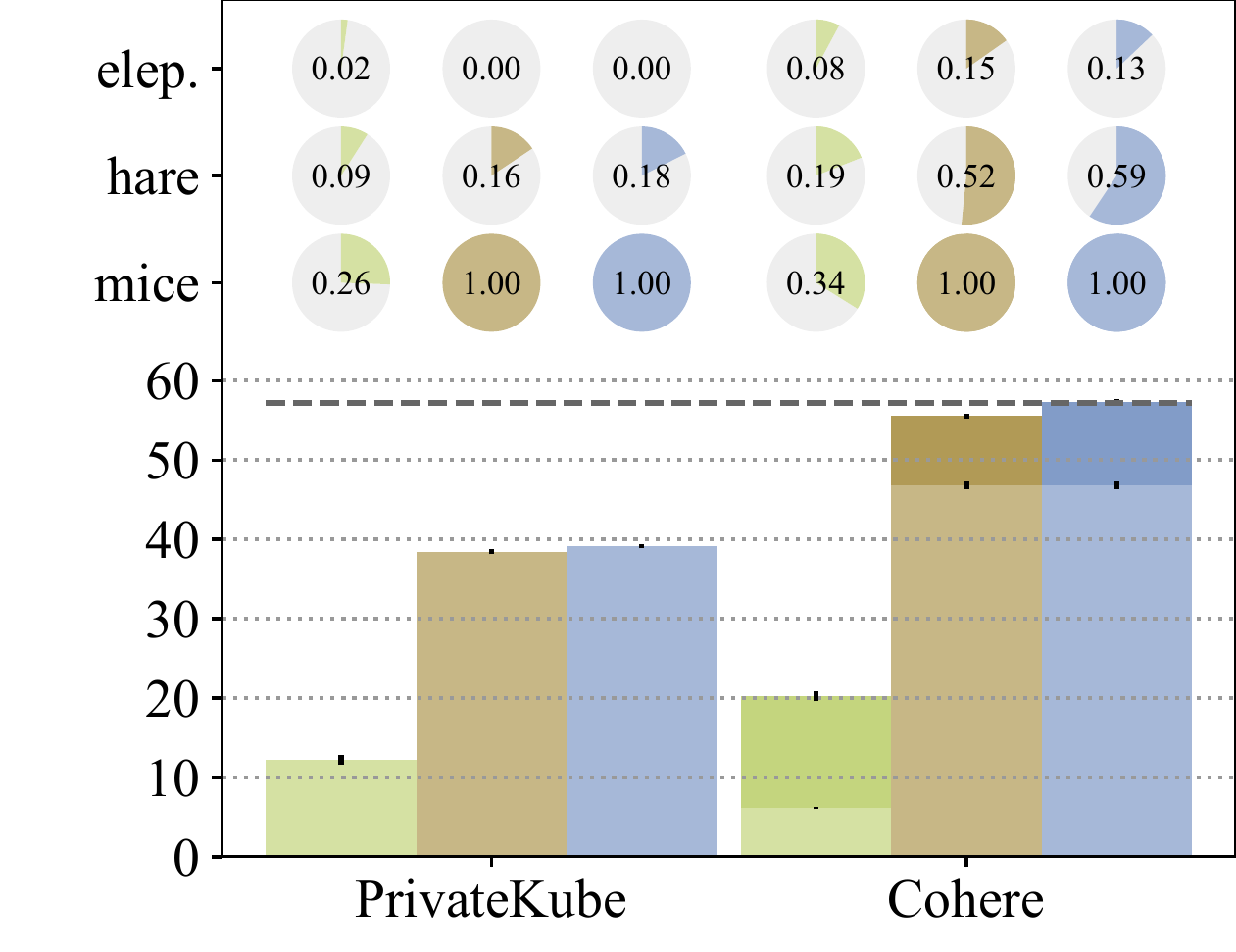}
    \end{subfigure}
    \begin{subfigure}[t]{0.24\textwidth}
        \includegraphics[width=1.1\linewidth]{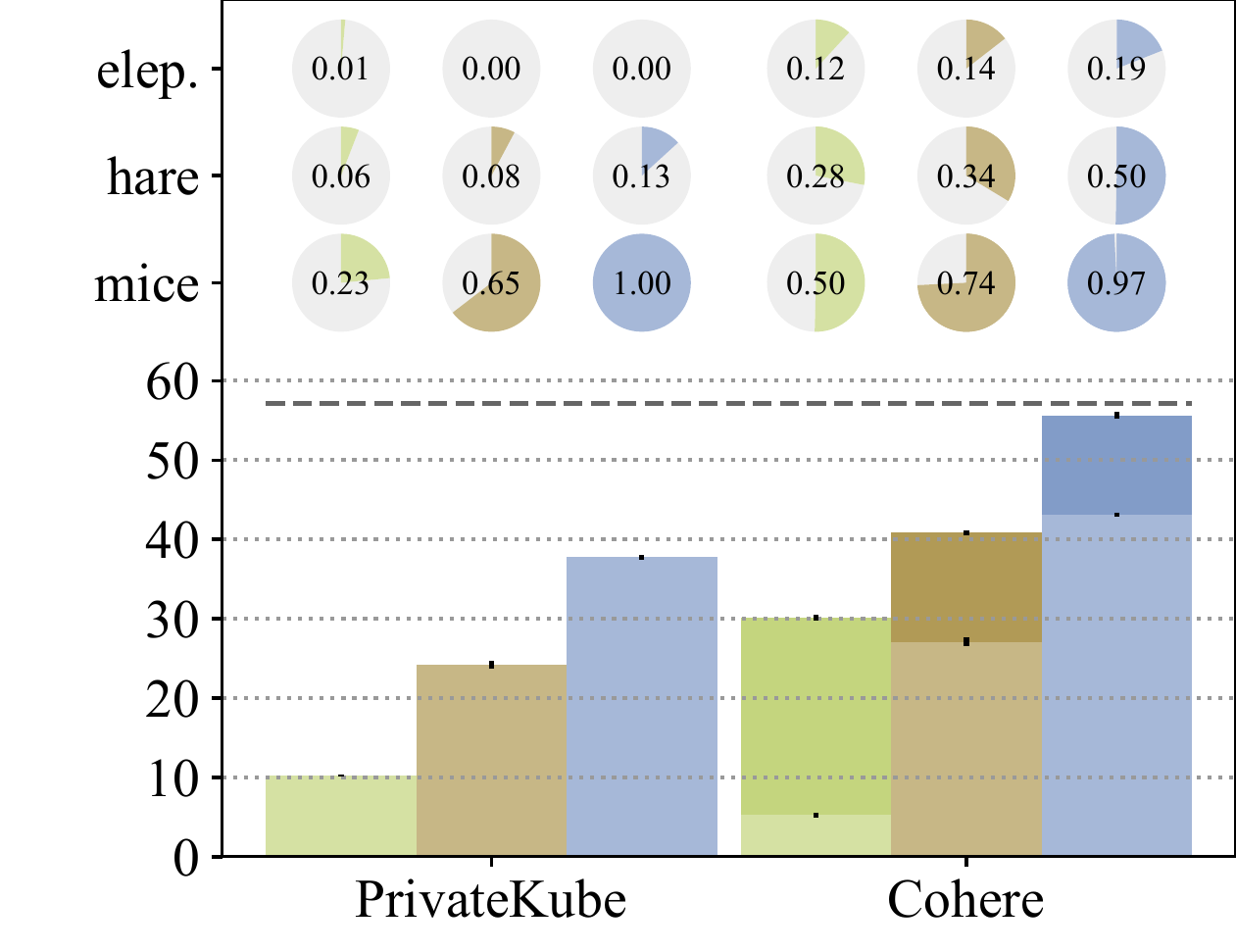}
    \end{subfigure}

    \vspace{0.2cm}

    \begin{subfigure}[t]{0.24\textwidth}
        \includegraphics[width=1.1\linewidth]{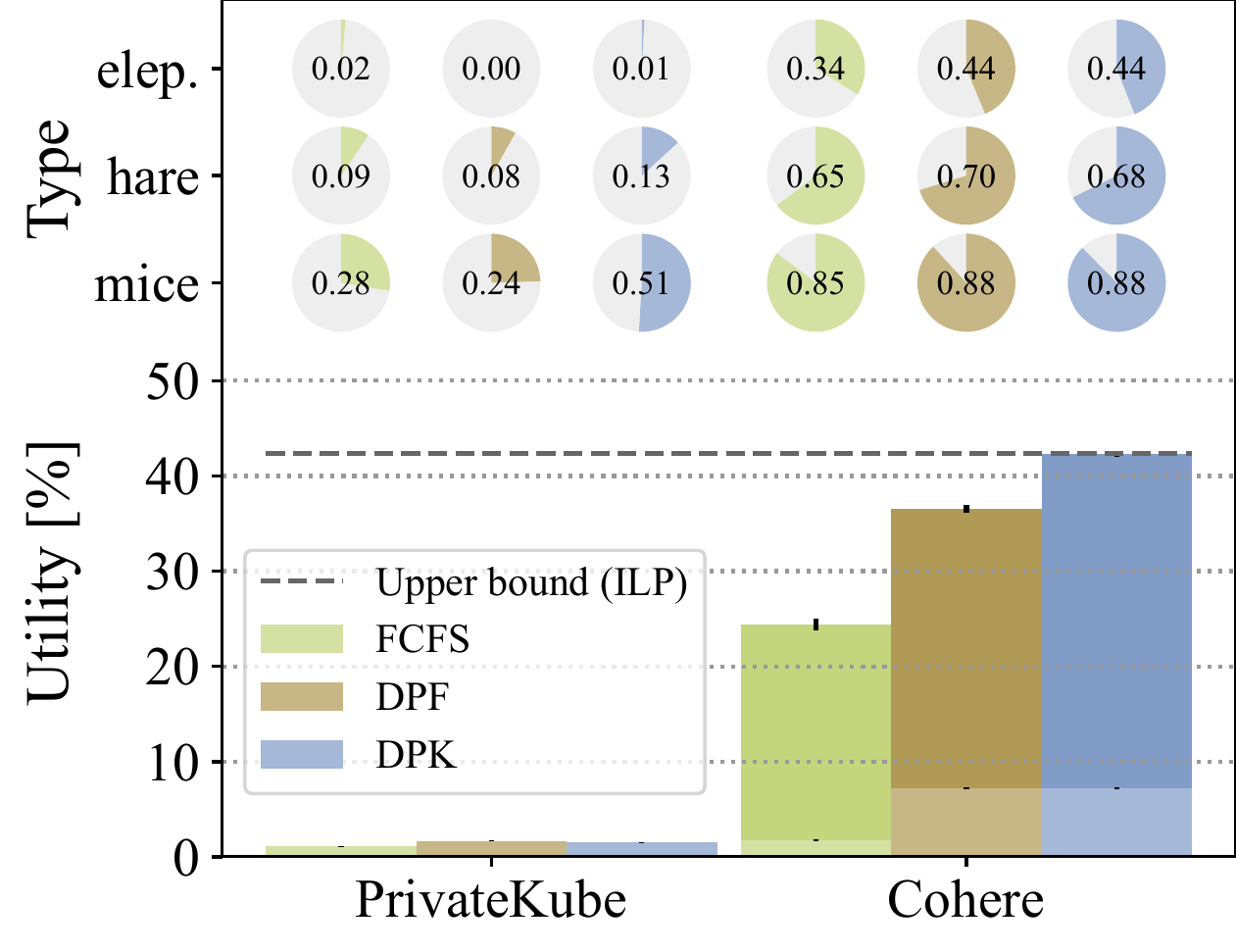}
        \caption{\gls{w:count}}
        \label{fig:w:gm}
    \end{subfigure}
    \begin{subfigure}[t]{0.24\textwidth}
        \includegraphics[width=1.1\linewidth]{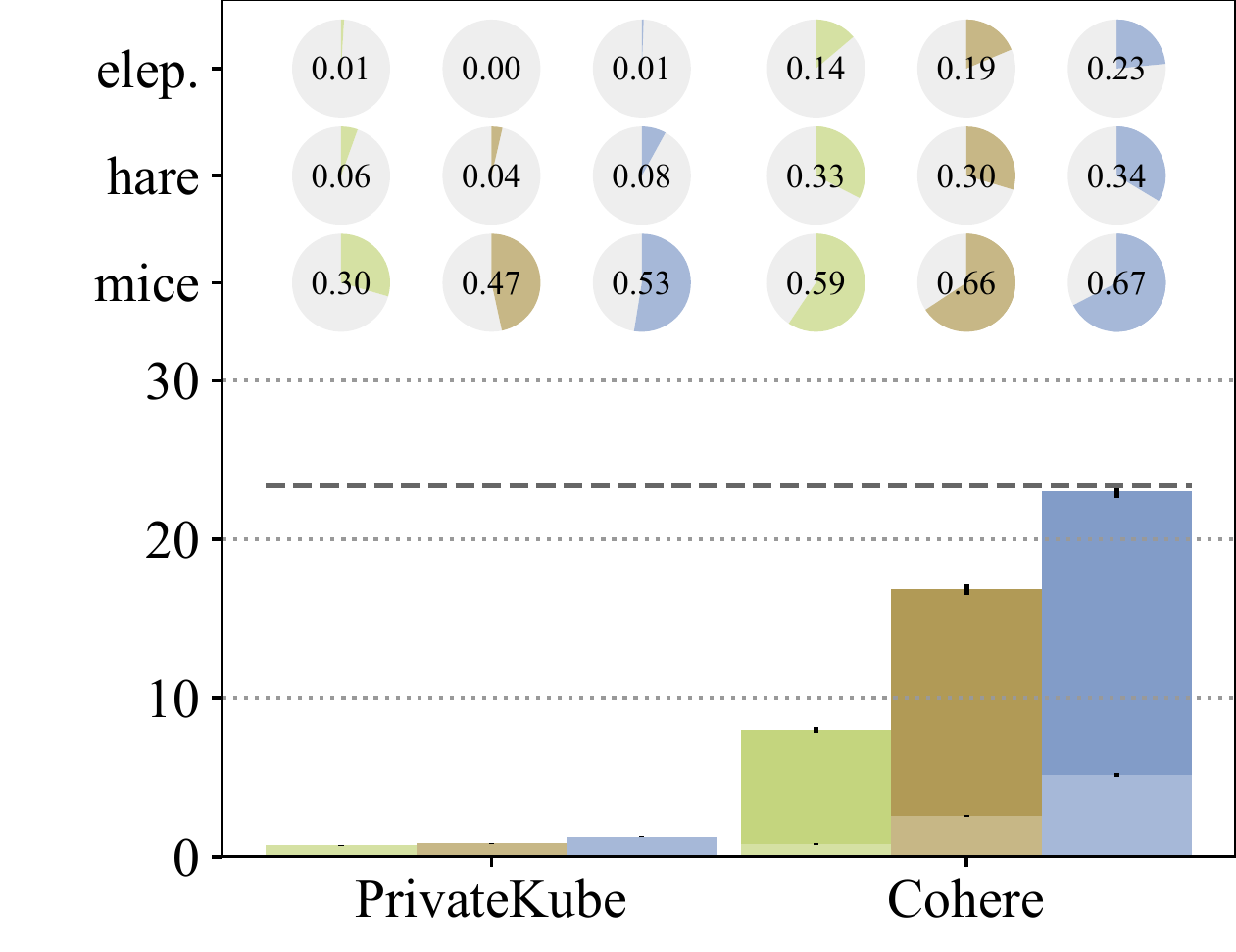}
        \caption{\gls{w:fund}}
        \label{fig:w:mix}
    \end{subfigure}
    \begin{subfigure}[t]{0.24\textwidth}
        \includegraphics[width=1.1\linewidth]{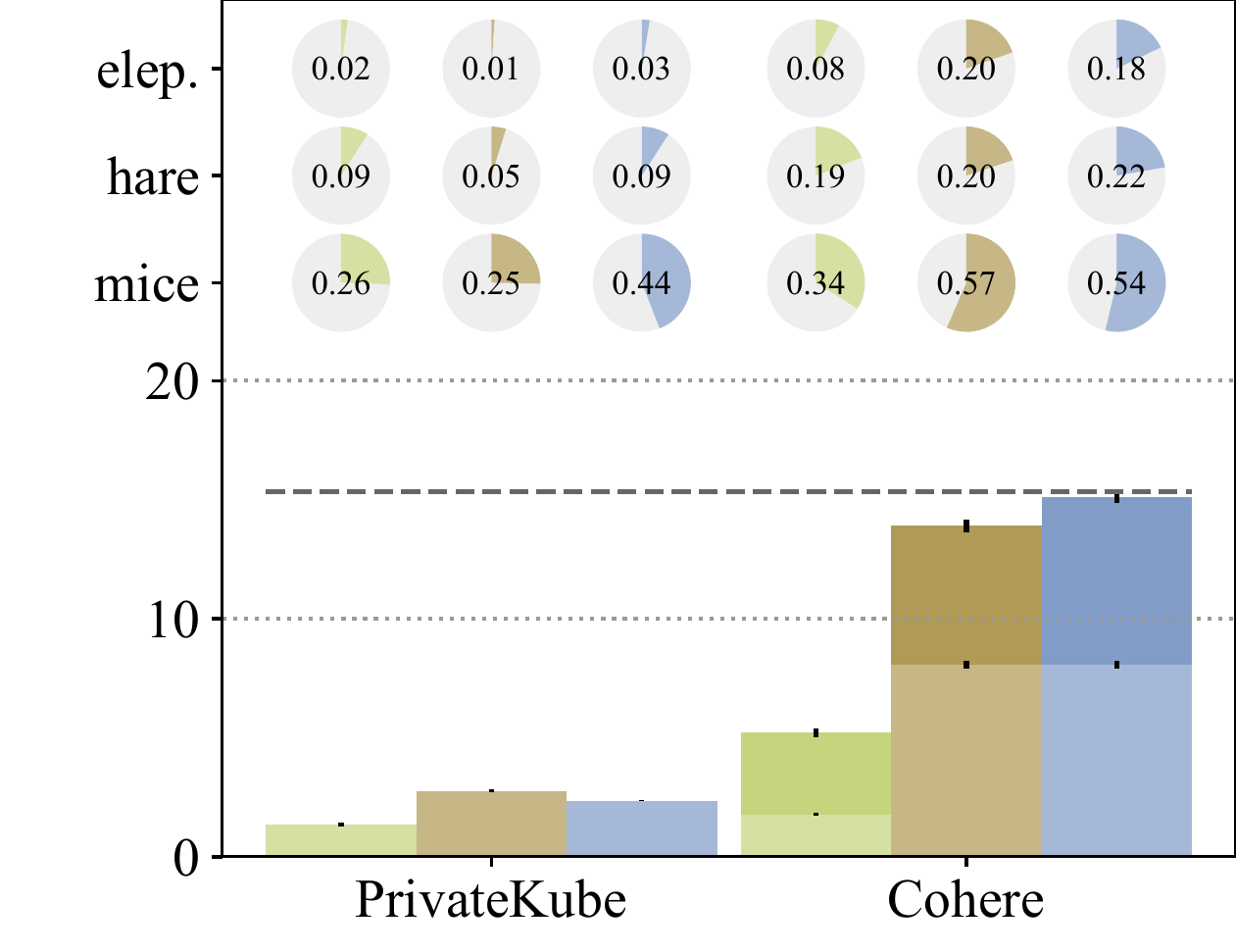}
        \caption{\gls{w:ml}}
        \label{fig:w:ml}
    \end{subfigure}
    \begin{subfigure}[t]{0.24\textwidth}
        \includegraphics[width=1.1\linewidth]{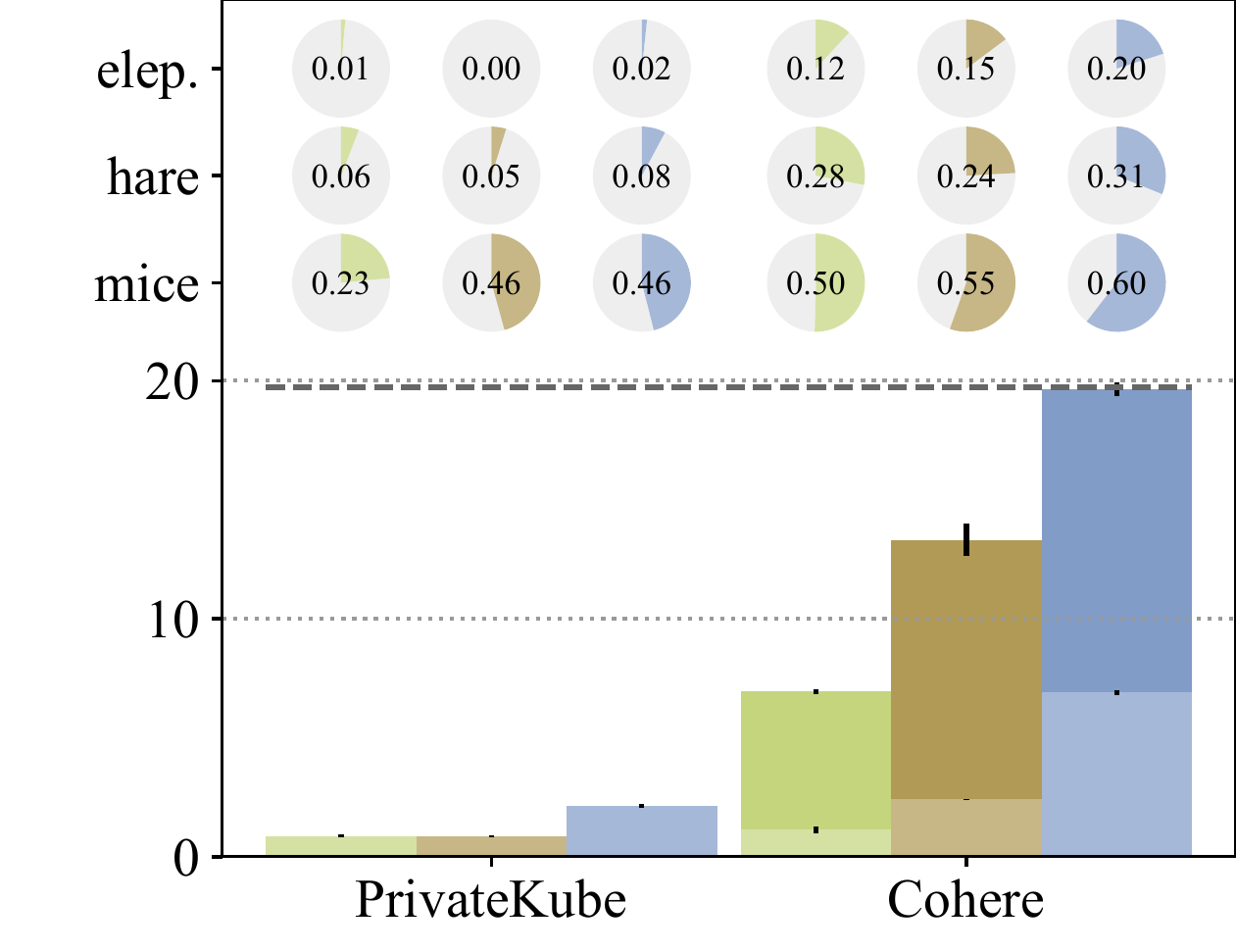}
        \caption{\gls{w:all}}
        \label{fig:w:all}
    \end{subfigure}
    \vspace{-2pt}
    \caption{
    Evaluation of \oursystem compared to \gls{privatekube} on four allocation approaches on our four workloads.
    The performance of \oursystem with \glspl{pa} is shown as a darker shade.}
\end{figure*}

\subsecspacingtop
\subsection{Evaluation Results}
\subsecspacingbot
\noindent
We compare \oursystem against \gls{privatekube}~\cite{Luo2021-privacysched, Tholoniat2022-dppacking}, which currently stands as the state-of-the-art solution for \gls{dp} resource management across multiple workloads.
First, we highlight the benefits of subsampling users as opposed to conducting a privacy analysis based on parallel composition at the individual user level, as used in \gls{privatekube}.
Subsequently, we show the advantages of implementing our new budget-unlocking strategy within a continuously running system with clean query semantics.
Finally, we compare \oursystem against \gls{privatekube}~\cite{Luo2021-privacysched, Tholoniat2022-dppacking} across a wide range of mixed workloads.
We evaluate both systems with different allocation algorithms: a greedy \gls{fcfs} baseline and two heuristic algorithms: (weighted) \gls{dpf}~\cite{Tholoniat2022-dppacking} and \gls{dpk}~\cite{Luo2021-privacysched, Tholoniat2022-dppacking}.
For \gls{privatekube}, we remove \gls{pa} information from requests, as it does not support them.
Since \gls{privatekube} requires requests to specify a concrete set of blocks (i.e., users), we follow the approach in their evaluation~\cite{Luo2021-privacysched, Tholoniat2022-dppacking} and select either all recent blocks (i.e., last 12 weeks) or uniformly sample from this population when smaller percentages are requested.
Note that a request that is accepted will be allocated (in expectation) the same number of users in \oursystem and \gls{privatekube}, ensuring a fair comparison.

\fakeparagraph{Benefits of Subsampling}
We show the benefits of privacy amplification via subsampling by considering variants of the workloads mentioned above.
We replace the equal split of 25\% and 100\% of users being requested with a single percentage varying from 5\% to 85\%.
\Gls{privatekube} utilizes parallel composition on the user level (\gls{upc}), i.e., they account for the privacy cost of the request only on the selected users.
Subsampling, by contrast, achieves lower overall privacy costs by distributing (significantly reduced) privacy costs across all users in the population (\cfref{sec:dp:sub}).
\Cref{fig:sub} presents the results for the \gls{w:count} and \gls{w:fund} workloads, with allocations determined via the \gls{dpk} algorithm.
Due to space constraints, we omit \gls{w:ml} and \gls{w:all} but note that the observed behavior remains consistent.
The relative improvement is highest when the percentage of expected users is low, but the Poisson subsampling outperforms \gls{upc} consistently across all percentages of expected users for both request volume and utility.
For the 25\% setting we will use in the remainder of the evaluation, we see improvements ranging from $1.17$x to $1.51$x in allocated requests and from $4.44$x to $4.98$x improvement in utility across the workloads (\Cref{fig:sub}).
This shows that subsampling's theoretical benefits~\cite{Zhu2019-rdpsample} translate into practical improvements, highlighting the importance of native subsampling support for effective privacy accounting.

\fakeparagraph{Benefits of our Budget Unlocking}
In \oursystem, we employ budget unlocking to impose guarantees on the availability of budget in each round, preventing a single set of requests from consuming the entire budget available in the sliding window.
The obvious strategy of unlocking $\frac{1}{12}$ of a block's budget for each round it is active does not compose well with subsampling:
requests could never use more than $\frac{1}{12}$ of the budget as they are limited by the budget of the most recent active block.
Our novel unlocking strategy, as described in~\Cref{sec:dp:cont}, helps to alleviate this issue by relaxing the available budget per round by a factor $\Delta$.
In \Cref{fig:unlock}, we show the effect of different $\Delta$s for the \gls{gm} workload, as it is the most heavily partitioned one and \glspl{pa} introduce greater variability in the allocated utility for different partitions.
We consider naive unlocking ($\Delta$ = 0), $\Delta=0.4$ and $\Delta$ = 0.8 and also show a no-unlocking baseline,
where we observe a large imbalance across rounds because the budget of the active set is consumed immediately in the first round.
Introducing some slack results in a total utility of 42\% for $\Delta$ = 0.4, compared to 25\% observed for $\Delta$ = 0.
However, higher slack, such as $\Delta$ = 0.8, causes excessive fluctuations in utility and thus violates fairness across rounds.
We choose $\Delta$ = 0.4 for the remainder of the evaluation, as it strikes a good balance between ensuring stability and maximizing utility across all workloads.

\fakeparagraph{Comparison to \gls{privatekube}}
In the remainder of the evaluation, we will focus on comparing \oursystem with \gls{privatekube} across all different workloads.
In \Cref{fig:w:gm,fig:w:mix,fig:w:ml,fig:w:all}, we present the fraction of allocated requests and the overall utility achieved by \gls{privatekube} and \oursystem using each of the three allocation algorithms (\gls{fcfs}, \gls{dpf}, \gls{dpk}).
In addition to the performance, we also show the fraction of mice, hares, and elephants accepted in each allocation.
Note, that 100\% utility is not achievable due to privacy budget constraints.
Therefore, we use an \gls{ilp} solver to compute the maximum achievable performance, shown as the dashed line.

We observe that the \gls{dpf} and \gls{dpk} heuristics significantly outperform the \gls{fcfs} baseline across all workloads.
Note that, for the \gls{fcfs} algorithm, the \gls{upc} accounting approach (\gls{privatekube}) sometimes outperforms our Poisson subsampling when optimizing for request volume.
Subsampling's tighter analysis forces \gls{fcfs} to accept high-cost requests which would be rejected under \gls{upc}, harming total request volume.
More sophisticated allocation algorithms can benefit from the tighter analysis while avoiding this pitfall.
For example, the \gls{dpf} and \gls{dpk} heuristics tend to allocate more mice than \gls{fcfs} when optimizing for request volume and more elephants when optimizing for utility.
Overall, the \gls{dpk} heuristic achieves within 96\% and 98\% of optimal request volume and utility, respectively.

Across all workloads, \oursystem significantly outperforms the prior state-of-the-art \gls{privatekube}~\cite{Luo2021-privacysched} in both request volume and utility.
Note, the maximum allocatable requests and utility are higher for the first workload  (see \Cref{fig:w:gm}) compared to the other workloads, which is expected because the \gls{gm} workload encodes the ideal case for \oursystem, with a heavily partitioned workload and only efficient composition across the Gaussian mechanism.
Even without considering \glspl{pa} in the resource planning, \oursystem improves the prior approach with \gls{dpk} by 1.1--1.2x in request volume and 3.2--4.8x in utility due to the use of Poisson subsampling.
\Glspl{pa} further increase this improvement to 1.5--2.0x in request volume and 6.4--28x in utility due to the more fine-grained privacy analysis.
Our results highlight the importance of a fine-grained privacy analysis in DP resource management systems. %
Moreover, our evaluation shows that integrating a more efficient and more fine-grained privacy analysis is more influential than the choice of the allocation algorithm.

\fakeparagraph{Performance Benchmark}
Finally, we discuss the runtime and memory consumption of \oursystem's planner.
Note, that \oursystem's planning system is not on the critical path for daily system usage, and that investing (usually abundant) computational resources for better management of the very scarce privacy budget is an attractive trade-off.
We find that it is feasible to solve the optimized version of the \gls{ilp} problem formulation across all our workload instances.
The \gls{ilp} takes an average of 48 minutes in total and a maximum memory consumption of $1.3$GB.
However, predicting the complexity of solving an \gls{ilp} instance is non-trivial and depends on a variety of factors.
The heuristic algorithms provide a good tradeoff in terms of performance and runtime.
With our optimizations, the total runtime of simulating 40 rounds for \gls{dpk} is only around 3.5 minutes without \glspl{pa}.
With \glspl{pa}, the runtime increases to 5.3 minutes for \Gls{w:count} and 22 minutes for \Gls{w:all}.
This is expected, as more complex workloads also increase the complexity of the optimization problem.
The requirement of \glspl{pa} to track multiple budgets per \gls{group} results in a somewhat higher maximum memory consumption of around 550MB for all workloads, compared to the 85MB required without \glspl{pa}.

\vspace{-2pt}

\subsecspacingtop
\section{Related Work}
\subsecspacingbot
\label{sec:relatedwork}

\fakeparagraph{Privacy Management for \gls{dp}}
PrivateKube~\cite{Luo2021-privacysched, Tholoniat2022-dppacking} is the most similar to our work, as it manages privacy resources for multiple ML pipelines.
However, its design falls short of addressing broader organizational needs, since it ties privacy management to hardware resource scheduling rather than providing a central privacy management layer.
PrivateKube prioritizes fairness by adapting the DRF~\cite{Ghodsi2011-drf} scheduling algorithm and follow-up work extended it to provide support for utility optimization~\cite{Tholoniat2022-dppacking}.
In comparison, our system focuses on improving the underlying privacy analysis by natively supporting subsampling and providing more granular resource allocation with partitioning attributes.
In addition, our user rotation provides clear query semantics and removes a source of bias compared to PrivateKube.

Sage~\cite{Lecuyer2019-sage} introduced block composition on user streams but only supports event-level \gls{dp} and relies on inefficient \gls{adp} composition.
Additionally, Sage does not cover resource allocation.
The planner by Li et al.~\cite{Li2023-dplanner} also uses \gls{adp} composition but considers user-to-application importance in the allocation.
While promising for some ML workloads, its general applicability is unclear, as most non-ML workloads do not have a notion of user importance, and the resulting non-uniform budget consumption requires specialized DP algorithms~\cite{Boenisch2022-dpindpate, Boenisch2023-dpsgdind}.
Some \gls{dp} platforms~\cite{Gaboardi2016-dppsi, Rogers2021-dplinkedin, Immuta2021-web, Zhang2023-dprovdb} provide limited budget management APIs but they lack support for resource allocation.

\fakeparagraph{Resource Allocation}
Previous research explored heuristic-based~\cite{Ghodsi2011-drf, Boutin2014-resheu, Gias2019-resheur, Yu2019-resheuml, Cortez2017-resheu, Vavilapalli2013-yarn}, ML-based~\cite{Rzadca2020-resml, Qiu2020-resfirm, Yu2019-resheuml, Pimpley2021-resopt}, and solver-based~\cite{Newell2021-resras, Narayanan2021-pop, Curino2014-ressolv, Tumanov2016-ressolv, Garefalakis2018-ressolv, Grandl2014-tetris, Grandl2016-resalltru, Grandl2016-ressolv} methods for computational resource allocation.
Our system focuses on utility-oriented privacy budget allocation, emphasizing allocation quality over speed due to the non-replenishable nature of \gls{dp} budgets.
Moreover, in the context of \gls{rdp}, privacy budget allocation diverges from (multi) resource allocation problems in other domains because costs and capacities cannot be efficiently represented as a single numerical value.
Instead, \gls{rdp} introduces a unique OR-semantic based on a set of costs and capacities.

\vspace{-4pt}

\section{Conclusion}
\label{conclusion}
\noindent
\gls{dp} offers significant potential for protecting privacy-sensitive information, but deploying it in large-scale systems remains challenging.
Managing shared application state and allocating non-replenishable privacy resources requires careful planning and coordination.
As \gls{dp} transitions from theory to practice, these deployment considerations become more critical.
This work sheds light on these new system challenges and presents \oursystem, a system that streamlines the use of DP in large-scale systems.
Our more fine-grained privacy analysis enables more applications within a given privacy budget, expanding \gls{dp}'s applicability without compromising guarantees.
We hope this work will enhance the understanding of the many systems challenges in this space and generate further work considering the challenges involved in deploying \gls{dp} in practice.

\ifdefined\isnotanon
\section*{Acknowledgments}
\noindent
We thank Lukas Burkhalter for his invaluable help in the early stages of the project and the anonymous reviewers for their insightful input and feedback.
We would also like to acknowledge our sponsors for their generous support, including Meta, Google, and SNSF through an Ambizione Grant No. PZ00P2\_186050.

\else
\fi

\clearpage
\bibliographystyle{plain2} %
\interlinepenalty=10000 %
\bibliography{references, referencesplus}
\interlinepenalty=0

\appendices %

\begin{appendices}

\section{Budget Unlocking \oursystem}
\label{sec:appendix:unlock}
\noindent
This section analyzes the process of unlocking budgets for each \gls{blockid}.
Recall that we have a sliding window over \glspl{group} $G_i$, and each \gls{group} is subdivided into \glspl{blockid}.
To simplify, we consider only the case with a single \gls{blockid} per \gls{group}.
However, the unlocking process remains identical with multiple \glspl{blockid}, and analogous properties hold.
In addition, we assume $\epsilon$-\gls{dp}.
Later, we will expand budget unlocking to \gls{rdp}.
For each allocation $i$, we consider the last $K$ \glspl{group} (i.e., $G_{i-K+1}$ until $G_i$) and we assume $K$ is even.
Each request selects all currently active \glspl{group} and, upon allocation, assigns the same cost to every selected \gls{group}.
The total allocated budget in round $i$ is denoted as $c_i$.

The budget-unlocking process functions as follows:
During the initial $\frac{K}{2}$ rounds of each \gls{group}'s lifetime, a budget of $(1+\Delta)\frac{\epsilon}{K}$ is unlocked.
Subsequently, in the latter $\frac{K}{2}$ rounds, a budget of $(1-\Delta)\frac{\epsilon}{K}$ per round is unlocked, where $\Delta \in [0,1]$.
After $K$ rounds, the \gls{group} is retired.

\noindent This process now fulfills several desirable properties:

\noindent
\underline{P1: Maximum Budget}: We never exceed the overall budget of $\epsilon$ for each \gls{group}.

\noindent
\textit{Proof}:
Consider that the unlocked budget in the final round of the allocation for each \gls{group} equals: $\sum_{i=1}^{K/2} \left( (1+\Delta)\frac{\epsilon}{K} \right) + \sum_{i=\frac{K}{2}+1}^{K} \left( (1-\Delta)\frac{\epsilon}{K} \right) = \frac{K}{2} (1+\Delta)\frac{\epsilon}{K} + \frac{K}{2} (1-\Delta)\frac{\epsilon}{K} = \frac{K}{2} (2 + \Delta - \Delta) \frac{\epsilon}{K} = \epsilon$. $\square$

\noindent
\underline{P2.1: Fairness 1}: For every allocation, there is at least $(1-\Delta)\frac{\epsilon}{K}$ of the budget available.

\noindent
\textit{Proof}:
Before every allocation, we unlock either $(1-\Delta)\frac{\epsilon}{K}$ or $(1+\Delta)\frac{\epsilon}{K}$ of budget in each \gls{group}. $\square$

\noindent
\underline{P2.2: Fairness 2}:
We never allocate more than $(1+\Delta)\frac{\epsilon}{K}$ in a single allocation round.

\noindent
\textit{Proof}:
We have only $(1+\Delta)\frac{\epsilon}{K}$ unlocked budget for the newest \gls{group}.
As each allocation must adhere to the unlocked budget on every active \gls{group}, the available budget in each allocation round cannot exceed $(1+\Delta)\frac{\epsilon}{K}$. $\square$

The allocation in each round depends on the available requests and the allocation algorithm.
However, the budget-unlocking strategy should possess the flexibility to enable as many allocations as possible.
First, we discuss two possible allocation series, and then we show what determines the available budget in each round.

\noindent
(\underline{P3.1: Greedy Optimal Allocation})
If we consistently allocate the entire available budget, then the allocation pattern follows a periodic sequence.
Assuming, that we start with a set of $K$ active \glspl{group} without any budget consumed, then we can allocate $(1+\Delta)\frac{\epsilon}{K}$ in the first $\frac{K}{2}$ rounds, and $(1-\Delta)\frac{\epsilon}{K}$ in the later $\frac{K}{2}$ rounds.

\noindent
\textit{Proof:}
In the first $\frac{K}{2}$ rounds, we can allocate $(1+\Delta)\frac{\epsilon}{K}$ per round, as the unlocking step of the respective first \gls{group} is always the limiting factor, i.e., $G_1$ in round 1, $G_2$ in round 2, etc.
After these $\frac{K}{2}$ rounds, the group $G_1$, which was newly added to the active set in the first round, has all the available budget consumed, i.e., $\frac{K}{2} \cdot (1+\Delta)\frac{\epsilon}{K}$.
In round $\frac{K}{2} + k$ for $k=1 \dots \frac{K}{2}$, this \gls{group} still has all available budget consumed and subsequently unlocks additional $(1-\Delta)\frac{\epsilon}{K}$ budget.
Consequently, our allocation during these rounds is constrained to $(1-\Delta)\frac{\epsilon}{K}$, as by \underline{P2.1}, this is always available.
After these $\frac{K}{2}$ rounds, the cost of the first round exits the active window, allowing us to resume allocating $(1+\Delta)\frac{\epsilon}{K}$.
In essence, whenever the cost of a certain magnitude exits the active window, we can reallocate this cost, creating a recurring pattern. $\square$

\noindent
(\underline{P3.2: Balanced Optimal Allocation}) The unlocking strategy allows for balanced allocations of $\frac{\epsilon}{K}$ in each round.

\noindent
\textit{Proof:}
The respective initial $\frac{K}{2}$ \glspl{group} impose no restriction since we unlock $(1+\Delta)\frac{\epsilon}{K} \geq \frac{\epsilon}{K}$ per round.
For a \gls{group} active for $k \geq \frac{K}{2}$ rounds, the past consumed budget is $\frac{(k-1) \cdot \epsilon}{K}$, and the unlocked budget is:
$B_k = \left[\frac{K}{2}(1+\Delta) \frac{\epsilon}{K} \right] + \left[ (k - \frac{K}{2})(1-\Delta)\frac{\epsilon}{K} \right] =(k + (K-k) \Delta) \frac{\epsilon}{K}$.
To show that we can allocate $\frac{\epsilon}{K}$, we need to show that $B_k - \frac{(k-1) \cdot \epsilon}{K} \geq \frac{\epsilon}{K}$.
Simplifying, this becomes $\frac{(K-k) \Delta \epsilon}{K} + \frac{\epsilon}{K} \geq \frac{\epsilon}{K}$, which holds true as $K > 0$, $k \leq K$, $\Delta \in [0,1]$, and $\epsilon > 0$, ensuring the first term is non-negative. $\square$

\noindent
(\underline{P4: Available Budget}) Given the allocation of at least $(1-\Delta)\frac{\epsilon}{K}$ in every round, the available budget $b_i$ can be expressed as: $b_i = \min\left(\epsilon - \sum_{j=i-K+1}^{i-1} c_j, (1+\Delta)\frac{\epsilon}{K}\right)$.

\noindent
Before showing \underline{P4}, we show that when each allocation incurs a cost of at least $(1-\Delta)\frac{\epsilon}{K}$, it suffices to address the budget constraints of the newest and the oldest \gls{group}.

\noindent
\textit{Proof:}
We collapse the budget constraints of the newer half of the \glspl{group} into the constraint of the newest group, which allows $(1+\Delta)\frac{\epsilon}{K}$ per round.
Next, we demonstrate that the budget constraints of the older half of the \glspl{group} can be collapsed into the constraint of the oldest \gls{group} in the active window.
Within the older half of \glspl{group}, for simplicity denoted by $G_1, G_2, \dots, G_{K/2}$ (where $G_1$ is the oldest), the available budget for group $G_i$ is the difference between the total budget $\epsilon$, the locked budget, and the consumed budget.
This can be expressed as $\epsilon - \left[ (i-1) \cdot (1-\Delta)\frac{\epsilon}{K} \right] - \left[c_i + \dots + c_{K-1}\right]$, where $c_i$ corresponds to the allocation cost when \gls{group} $G_i$ was the newest \gls{group} in the active window.
As all $c_k \geq (1-\Delta)\frac{\epsilon}{K}$, the available budget in \gls{group} $G_1$ is the most restricted. $\square$

\noindent
\textit{Proof P4:}
As shown above, we only need to consider the budget constraints of the newest and the oldest \gls{group} to determine the overall available budget.
The newest \gls{group} allows an allocation of up to $(1+\Delta)\frac{\epsilon}{K}$ but no more, leading to the $min\left( \: \cdot \: , (1+\Delta)\frac{\epsilon}{K}\right)$ condition.
In the case of the oldest \gls{group}, no budget is locked. Therefore, the available budget is the difference between the total budget and the sum of the last $K-1$ allocations, represented as $\epsilon - \sum_{j=i-K+1}^{i-1} c_j$. $\square$

To extend the budget unlocking strategy to \gls{rdp}, we unlock the budget as we described for $\epsilon$-\gls{dp}, but for each $\alpha_a$ for $a \in \mathcal{A}$ individually, in steps of $(1\pm \Delta)\frac{\epsilon_{\alpha_a}}{K}$.
$\epsilon_{\alpha_a}$ is the budget for $\alpha_a$ and $\mathcal{A} = \{1, ..., A_{max}\}$ is the index set of tracked values of $\alpha_a$.
The described properties hold, however, slightly modified:
\begin{itemize}
    \item \underline{P1}: Budget $\epsilon_{\alpha_a}$ is not exceeded at least for a single value of $a \in \mathcal{A}$, which means the budget overall is not exceeded by \gls{rdp} semantics.
    \item \underline{P2.1}: For every allocation, we have at least $(1-\Delta)\frac{\epsilon_{\alpha_a}}{K}$ in budget available for some $a \in \mathcal{A}$ (we might have less available for other values of $a$).
    \item \underline{P2.2}: For every allocation, we allocate at most $(1+\Delta)\frac{\epsilon_{\alpha_a}}{K}$ in budget for some $a \in \mathcal{A}$ (we might allocate more for other values of $a$).
    \item \underline{P3.1}/\underline{P3.2}: Holds for some value of $\alpha_a$, i.e., the active $\alpha_a$.
    \item \underline{P4}: Holds for all $\alpha_a$, however, for some the available budget may be negative, i.e., inactive $\alpha_a$.
\end{itemize}

\section{Privacy Resource Allocation ILP}
\label{sec:appendix:ilp}

\noindent
We want to automatically compute an optimal assignment of \glspl{group} to requests which maximizes the utility while maintaining the privacy budgets for each \gls{blockid} (grouped into segments).
We define our optimization problem as an \gls{ilp}, i.e., a mathematical optimization problem, in which the objective function and the constraints are linear, and the variables are all integers.
We present an \gls{ilp} formulation that supports multiple allocation approaches for \gls{rdp} block composition.
In addition to addressing the optimization problem outlined in \Cref{sec:alloc}, this formulation can accommodate an alternative allocation approach where certain requests might require only a specific number of the active \glspl{group}, rather than all of them.
The optimization problem shares similarities with the 0/1 multiple multidimensional knapsack problem (0/1 MMKP)~\cite{Song2008-mmkp, Laabadi2018-mmkp}, where items have costs in multiple dimensions and must be (optionally) packed into one of several knapsacks with scalar capacities for each dimension.
In our case, requests represent items, \glspl{group} are the knapsacks, and segments are dimensions.
Our problem differs in two aspects.
First, assignments must be made to multiple knapsacks due to requests requiring multiple \glspl{group}.
Moreover, \gls{rdp} composition introduces a novel logical OR-semantic for segment capacities.
Instead of using scalar capacities for each dimension as in 0/1 MMKP, our \gls{ilp} tracks for each dimension a set of costs and capacities for different orders $\alpha_a$.
The semantic is that the budget constraint of a dimension is satisfied, if the budget constraint for a single order is satisfied.

\begin{figure}[t]
    \renewcommand\figurename{Listing}
    \begin{align*}
         & \underset{}{\text{max}} &  & \smashoperator{\sum_{i \in \mathcal{R}}} \ y_i \cdot W_i          &  &                                                                      \\
         & \text{s.t.}             &  & \smashoperator{\sum_{j \in \mathcal{N}}} \ x_{ij} \geq y_{i} \cdot D_i &  & \mscale{0.9}{[\forall i \in \mathcal{R}]}                                                                              \\
         &                         &  & \smashoperator{\sum_{x_{ij} \in \mathcal{X}(S_k)}}  \ (x_{ij}-z_{ka}) \cdot C_{i}^{(a)} \leq B_{k}^{(a)} &  & \mscale{0.9}{[\forall k \in \mathcal{S}][\forall a \in \mathcal{A}]}                                                                           \\
         &                         &  & \smashoperator{\sum_{a \in \mathcal{A}}} \ z_{ka} < |\mathcal{A}| &  & \mscale{0.9}{[\forall k \in \mathcal{S}]}                                                                       \\
         &                         &  & y_{i} \in \{0,1\}                   &  & \mscale{0.9}{[\forall i \in \mathcal{R}]}                            \\
         &                         &  & x_{ij} \in \{0,1\}                  &  & \mscale{0.9}{[\forall i \in \mathcal{R}][\forall j \in \mathcal{N}]} \\
         &                         &  & z_{ka} \in \{0,1\}                  &  & \mscale{0.9}{[\forall k \in \mathcal{S}][\forall a \in \mathcal{A}]} \\
    \end{align*}
    \vspace{-1cm}
    \caption{Integer Linear Program (ILP) formulation to find an optimal \gls{allocation}.}
    \label{lis:ilp}
\end{figure}

In \Cref{lis:ilp}, we provide a complete formulation of the ILP.
Below, we provide explanations for each part.

\fakeparagraph{Notation}
Let $\mathcal{R} \coloneqq \{1, \; \ldots, \; R_{max}\}$, $\mathcal{N} \coloneqq \{1, \; \ldots, \; N_{max}\}$ and $\mathcal{S} \coloneqq \{1, \; \ldots, \; S_{max}\}$ denote index sets of the requests, \glspl{group} and segments respectively.
A request $R_i$ for $i \in \mathcal{R}$ is a tuple $(D_i,\Phi_i,\mathbf{C}_i,W_i)$ consisting of a \emph{data requirement} requesting $D_i \in \mathbb{N}$ \glspl{group}\!\footnote{
Note, when using Poisson subsampling, we set the data requirements such that all active \glspl{group} are required for a request to be fulfilled, i.e., $D_i = N_{max}$.
Technically, this problem corresponds to the 0/1 multidimensional knapsack problem (0/1 MKP) with the same logical OR-semantic for capacities (\cfref{sec:alloc}).
} filtered by a propositional formula $\Phi_i$ over the \glspl{pa},
a \emph{budget requirement} expressed as a vector $\mathbf{C}_i \in \mathbb{R}_{\geq 0}^{|\mathcal{A}|}$ of \gls{rdp} costs for different orders $\alpha_a$ (for $a \in \mathcal{A} \coloneqq \{1, \; \ldots, \; A_{max}\}$),
and a \emph{weight}\footnote{Weights are determined either by utility or set  to $W_i = 1  \ \forall i \in \mathcal{R}$ when maximizing the number of requests.} $W_i \in \mathbb{N}$.
\Glspl{group} are uniquely identified by their global ID $j \in \mathcal{N}$, thus we do not differentiate between \glspl{group} and their IDs (not to be confused with \glspl{blockid}). The notion of segments is introduced in \Cref{sec:alloc}.

\fakeparagraph{Decision Variables \& Objective}
We introduce two main types of decision variables.
First,  $y_i \in \{0, 1\}$ for $i \in \mathcal{R}$, where $y_i = 1$ means the request $R_i$ is accepted, and $y_i = 0$ means the request has been rejected.
Second, the decision variables $x_{ij} \in \{0, 1\}$ for $i \in \mathcal{R}$, $j \in \mathcal{N}$ indicate whether or not \gls{group} $j$ is allocated to request $R_i$.
The \emph{objective} is to maximize the sum of weights of the accepted requests.

\fakeparagraph{Constraints}
Finally, there are two main types of constraints: \emph{data constraints} that ensure that an accepted request is assigned sufficient \glspl{group}, and \emph{budget constraints} that enforce correct privacy budget accounting given the \glspl{pa} and budget requirement of each request.
The \emph{data constraints} (line 2 in \Cref{lis:ilp}) ensure that request $R_i$ is only allocated if the number of assigned \glspl{group} is greater or equal to the demand $D_i$.
The \emph{budget constraints} are more complex, as they rely on the notion of segments.
$\mathcal{X}(S_k)$ indirectly encodes the propositional formulas $\Phi_i$.
The core idea is that the \gls{ilp} ensures that for each segment, there is at least one value of $\alpha$ (line 4) such that we do not use more $\epsilon^{(rdp)}$ (line 3) than the budget for that value of $\alpha$.

\end{appendices}

\section{Meta-Review}
\noindent
The following meta-review was prepared by the program committee for the 2024
IEEE Symposium on Security and Privacy (S\&P) as part of the review process as
detailed in the call for papers.

\subsection{Summary}
\noindent
This paper presents Cohere, a system for efficiently managing differential privacy in large-scale systems where coordination and planning is required to support applications with diverse workloads and privacy requirements.

\subsection{Scientific Contributions}
\begin{itemize}
\item Addresses a Long-Known Issue
\item Provides a Valuable Step Forward in an Established Field
\end{itemize}

\subsection{Reasons for Acceptance}

\begin{enumerate}
	\item The paper presents a complete framework and system for managing differential privacy applications in real-world scenarios. Handling applications with different privacy constraints is a known challenge for differential privacy.

	\item The paper provides a valuable step forward in an established field. Compared to existing solutions to manage privacy resources (e.g., PrivateKube), Cohere empirically claims significantly improved utility. This is achieved through several novel elements, including the combination of user rotation with standard differential privacy techniques such as partitioning and sampling.
\end{enumerate}

\end{document}